\documentclass{amsart}
\usepackage[all]{xy}
\usepackage[breaklinks=true]{hyperref}
\usepackage{graphics}
\usepackage{color}

\usepackage{rotating}
\usepackage{hyperref}

\begin{document}

	\theoremstyle{plain}
\newtheorem{theorem}{Theorem}[section]
\newtheorem{proposition}[theorem]{Proposition}
\newtheorem{corollary}[theorem]{Corollary}

\newtheorem{observation}[theorem]{Observation}

\newtheorem{theoremapp}{Theorem A.\!\!}
\newtheorem{propositionapp}{Proposition A.\!\!}

\newtheorem{corollaryapp}{Corollary A.\!\!}
\newtheorem{observationapp}{Observation A.\!\!}
\newtheorem{lemmaapp}{Lemma A.\!\!}

\theoremstyle{definition}
\newtheorem{definition}[theorem]{Definition}
\newtheorem{dfn}[theorem]{definition}
\newtheorem{remark}[theorem]{Remark}
\newtheorem{example}[theorem]{Example}
\newtheorem{examples}[theorem]{Examples}

\newtheorem{definitionapp}{Definition A.\!\!}
\newtheorem{dfnapp}{definitionapp}
\newtheorem{remarkapp}{Remark A.\!\!}
\newtheorem{exampleapp}{Example A.\!\!}

\newdir^{ (}{{}*!/-5pt/@^{(}}

\def\proof {{\it Proof.}\hspace{7pt}}

\def\endofproof{\hfill{$\square$}\\}
\def\dfn{Definition}
\def\rmk{Remark}
\def\Cech{{{\v C}ech}}

\def\Loo{{$L_\infty$}}

\pagestyle{myheadings}

\title[SUPER LIE $n$-ALGEBRAS AND SUPER $p$-BRANES]{SUPER LIE $n$-ALGEBRA EXTENSIONS, HIGHER WZW MODELS
 AND SUPER $p$-BRANES WITH TENSOR MULTIPLET FIELDS}

 \author{Domenico Fiorenza}
\address{Dipartimento di Matematica,``La Sapienza'' Universit\`a di Roma,\newline P.le Aldo Moro, 5, Roma, Italy \newline{\tt fiorenza@mat.uniroma1.it}}
\author{Hisham Sati}
\address{Department of Mathematics,  University of Pittsburgh, \newline  Pittsburgh, PA 15260 \newline{\tt hsati@pitt.edu}}
\author{Urs Schreiber}
\address{Charles University Institute of Mathematics, \newline Sokolovsk\'a 83, 186 75 Praha 8, Czech Republic \newline{\tt urs.schreiber@googlemail.com}}

\maketitle

\begin{abstract}
We formalize higher dimensional and higher gauge WZW-type sigma-model local
  prequantum field theory, and discuss its rationalized/perturbative
  description in (super-)Lie $n$-algebra homotopy theory
  (the true home of the ``FDA''-language used in the supergravity literature).
  We show generally how the intersection laws
  for such higher WZW-type $\sigma$-model branes (open brane ending on background brane)
  are encoded precisely in (super-)$L_\infty$-extension theory and how the resulting
  ``extended \mbox{(super-)}spacetimes'' formalize spacetimes containing
  $\sigma$-model brane condensates. As an application
  we prove in Lie $n$-algebra homotopy theory that the
  complete super $p$-brane spectrum of superstring/M-theory is realized this way,
  including the pure sigma-model branes (the ``old brane scan'')
  but also the branes with tensor multiplet worldvolume fields,
  notably the D-branes and the M5-brane. For instance the degree-0
  piece of the higher symmetry algebra of 11-dimensional spacetime with an
  M2-brane condensate turns out to be the ``M-theory super Lie algebra''.
  We also observe that in this
  formulation there is a simple formal proof of the fact that
  type IIA spacetime with a D0-brane condensate is
  the 11-dimensional sugra/M-theory spacetime, and of (prequantum) S-duality
  for type IIB string theory. Finally we give the non-perturbative description of all
  this by higher WZW-type $\sigma$-models on higher super-orbispaces
  with higher WZW terms in stacky differential cohomology.
\end{abstract}

\def\wedgebullet {{\mbox{$\bigwedge^\bullet$}}}

\def\gg {\mathfrak{g}}
\def\hh {\mathfrak{h}}
\def\uu {\mathfrak{u}}

\newcommand{\attn}[1]{\begin{center}\framebox{\begin{minipage}{8cm}#1
\end{minipage}}\end{center}}

\renewcommand{\(}{\begin{equation}}
\renewcommand{\)}{\end{equation}}
\newcommand{\bea}{\begin{eqnarray}}
\newcommand{\eea}{\end{eqnarray}}
\newcommand{\R}{{\mathbb R}}
\newcommand{\C}{{\mathbb C}}
\newcommand{\Z}{{\mathbb Z}}
\newcommand{\Q}{{\mathbb Q}}
\newcommand{\G}{{\mathcal G}}

\def\proof {{Proof.}\hspace{7pt}}

\def\endofproof {\hfill{$\Box$}\\}



\section{Introduction: Traditional WZW and the need for higher WZW}

For $G$ be a simple Lie group, write $\mathfrak{g}$ for its semisimple Lie algebra.
The Killing form invariant polynomial $\langle -,-\rangle : \mathrm{Sym}^2 \mathfrak{g} \to \mathbb{R}$ induces the canonical Lie algebra 3-cocycle
$$
  \mu := \langle-,[-,-]\rangle : \mathrm{Alt}^3(\mathfrak{g}) \to \mathbb{R}
$$
which by left-translation along the group defines the canonical
closed and left-invariant 3-form
$$
  \langle \theta \wedge [\theta\wedge \theta]\rangle
  \in
  \Omega^3_{\mathrm{cl}, \mathrm{L}}(G)
  \,,
$$
where $\theta \in \Omega^1_{\mathrm{flat}}(G,\mathfrak{g})$ is the
canonical \emph{Maurer-Cartan form} on  $G$.
What is called the \emph{Wess-Zumino-Witten sigma-model} induced by this data
(see for instance \cite{CaseStudy} for a decent review) is
the prequantum field theory given by an action functional, which
to a smooth map $\Sigma_2 \to G$ out of a
closed oriented smooth 2-manifold assigns the product of the standard
exponentiated kinetic action with an exponentiated ``surface holonomy''
of a 2-form connection whose curvature 3-form is
$\langle \theta \wedge [\theta\wedge \theta]\rangle$.

\smallskip
In the special case that $\phi : \Sigma_2 \to G$ happens to factor through a contractible open subset $U$ of $G$ --
notably in the \emph{perturbative expansion} about maps constant on a point --
the Poincar{\'e} lemma implies that one can find a potential 2-form $B \in \Omega^2(U)$
with $d B = \langle \theta \wedge [\theta\wedge \theta]\rangle|_{U}$ and with this
perturbative perspective understood
one may take the action functional to be simply of the naive form that is often considered
in the literature:
$$
 \exp(i S_{\mathrm{WZW}})
 :=
 \exp\left(i \int_{\Sigma^2} \mathcal{L}_{\mathrm{WZW}}\right)
  \;:\;
  \phi \mapsto \exp\left(2\pi i \int_{\Sigma_2} \phi^\ast B\right)
  \,.
$$

\smallskip
There are plenty of hints and some known examples which point to
the fact that this construction
of the standard WZW model is just one in a large class of examples of
higher dimensional boundary local (pre-)quantum field theories
\cite{lpqft, Nuiten} which generalize
traditional WZW theory in two ways:
\begin{enumerate}
  \item the cocycle $\mu$ is allowed to be of arbitrary degree;
  \item the Lie algebra $\mathfrak{g}$ is allowed to be a
   (super-)\emph{Lie $n$-algebra} for $n \geq 1$ ($L_\infty$-algebra).
\end{enumerate}
One famous class of examples of the first point are the Green-Schwarz
type action functionals for the super $p$-branes of string/M-theory \cite{AETW}.
These are the higher dimensional analog of the action functional for the
superstring that was first given in \cite{GreenSchwarz} and then recognized as
a super WZW-model in \cite{HM}, induced from an exceptional 3-cocycle
on super-Minkowski spacetime of bosonic dimension 10,
regarded a super-translation Lie algebra.
Thess higher dimensional Green-Schwarz type $\sigma$-model
action functionals are accordingly induced by higher
exceptional super-Lie algebra cocycles on super-Minkowski
spacetime, regarded as a super-translation Lie algebra.
Remarkably, while ordinary Minkowski spacetime is cohomologically
fairly uninteresting, super-Minkowski spacetime has a
finite number of \emph{exceptional} super-cohomology classes.
The higher dimensional WZW models induced by the corresponding
higher exceptional cocycles
account precisely for the $\sigma$-models of those super-$p$-branes in
string/M-theory which are pure $\sigma$-models, in that they do not
carry (higher) gauge fields (``tensor multiplets'') on their worldvolume,
a fact known as ``the old brane scan'' \cite{AETW}. This includes, for instance, the
heterotic superstring and the M2-brane, but excludes the D-branes and the M5-brane.

\smallskip
However, as we discuss below in
section \ref{SuperBranesAndTheirIntersectionLaws}, this restriction
to pure $\sigma$-model branes without ``tensor multiplet'' fields on
their worldvolume is due
to the restriction to ordinary super Lie algebras, hence to super Lie $n$-algebras
for just $n = 1$. If one allows genuinely higher WZW models which are given
by higher cocycles on Lie $n$-algebras for higher $n$, then \emph{all} the
super $p$-branes of string/M-theory are described by higher WZW $\sigma$-models.
This is an incarnation of the general fact that in higher differential geometry, in the sense
of \cite{FSS,cohesive}, the distinction between $\sigma$-models and (higher) gauge theory
disappears, as (higher) gauge theories are equivalently $\sigma$-models
whose target space is a smooth higher moduli \emph{stack}, infinitesimally
approximated by a Lie $n$-algebra for higher $n$.

\smallskip
This general phenomenon is particularly interesting for the M5-brane
(see for instance the Introduction of \cite{7d} for plenty of pointers to the
literature on this). According to
the higher Chern-Simons-theoretic formulation of $\mathrm{AdS}_7/\mathrm{CFT}_6$ in
\cite{Witten}, the 6-dimensional $(2,0)$-superconformal worldvolume
theory of the M5-brane is related to the 7-dimensional Chern-Simons term
in 11-dimensional supergravity compactified on a 4-sphere in direct analogy
to the famous relation
of 2d WZW theory to the 3d-Chern-Simons theory
controled by the cocycle $\mu$ (see \cite{CaseStudy} for a review).
In previous work we have discussed the
bosonic nonabelian (quantum corrected) component of this 7d Chern-Simons theory
as a higher gauge local prequantum field theory \cite{7d, CField}; the discussion here
provides the fermionic terms and the formalization
of the 6d WZW-type theory induced from a (flat) 7-dimensional Chern-Simons theory.

\smallskip
Up to the last section in this paper we discuss general aspects
and examples of higher WZW-type
sigma-models in the rational/perturbative approximation,
where only the curvature $n$-form matters
while its lift to a genuine cocycle in differential cohomology is ignored.
However, in order to define already the traditional WZW action
functional in a sensible way on \emph{all}
maps to $G$, one needs a more global description of the WZW term $\mathcal{L}_{\mathrm{WZW}}$.
Since \cite{Ga, FW}, this is understood to be a circle 2-connection/bundle gerbe/Deligne 3-cocycle whose curvature 3-form is
$\langle \theta \wedge [\theta\wedge \theta]\rangle$, hence a
\emph{higher prequantization} \cite{hgp} of the curvature 3-form,
which, following \cite{FSS,cohesive} we write as a lift of maps
of smooth higher stacks
$$
  \xymatrix{
    && \mathbf{B}^2 U(1)_{\mathrm{conn}}
	\ar[d]^{H_{(-)}}
    \\
    G \ar@{-->}[urr]^-{\mathcal{L}_{\mathrm{WZW}}}
	\ar[rr]_{\langle \theta \wedge [\theta\wedge \theta]\rangle}
	&&
	\Omega^3_{\mathrm{cl}} \;,
  }
$$
where $\mathbf{B}^2 U(1)_{\mathrm{conn}}$ denotes the smooth 2-stack of
smooth circle 2-connections.
Then for $\phi : \Sigma_2 \to G$ a smooth map from a closed
oriented 2-manifold to $G$,
the globally defined value of the action functional is
the corresponding \emph{surface holonomy} expressed as the composite
$$
  \exp(i S_{\mathrm{WZW}})
  \;:=\;
  \exp\left(
    2 \pi i \int_{\Sigma_2} [(-),\mathcal{L}_{\mathrm{WZW}}]
  \right)
  \;,\;
  $$
  $$
  \xymatrix{
    [\Sigma,G]
	\ar[rr]^-{[\Sigma, \mathcal{L}_{\mathrm{WZW}}]}
	&&
	[\Sigma,\mathbf{B}^2 U(1)_{\mathrm{conn}}]
	\ar[rrr]^-{\exp(2 \pi i \int_{\Sigma_2})(-) }
	&&&
	U(1)
  }
  \,,
$$
of the functorial mapping stack construction followed by a stacky refinement
of fiber integration in differential cohomology,
as discussed in \cite{higher-extended, StackyPerspective}.

\smallskip
Towards the end, in section \ref{Outlook} we demonstrate a
general universal construction of
such non-perturbative refinements of all the local higher WZW terms considered in the
main text. We show how these are in a precise
sense boundary local prequantum field theories for flat higher
Chern-Simons type local prequantum field theories as explained in \cite{hgp, lpqft}
(which is in line with the Chern-Simons theoretic holography in \cite{Witten}).
Therefore we know in principle how to
quantize them non-perturbatively in generalized cohomology,
namely along the lines of \cite{Nuiten}. This, however, is to be discussed elsewhere.

\section{Lie $n$-algebraic formulation of perturbative higher WZW}

We start with the traditional WZW model and show how in
this example we may usefully reformulate its rationalized/perturbative aspects
in terms of Lie $n$-algebraic structures. Then we
naturally and seamlessly generalize to a definition of higher WZW-type $\sigma$-models.

\smallskip
We recall the notion of $L_\infty$-algebra valued
differential forms/connections
from \cite{SSS, FSS} to establish our notation. All the actual $L_\infty$-homotopy
theory that we need can be found discussed or referenced in \cite{frsII}.
Just for simplicity of exposition and since it is sufficient for the present discussion,
here we take all $L_\infty$-algebras to be of finite type, hence degreewise finite dimensional; see \cite{Pridham}
for the general discussion in terms of pro-objects.

\smallskip
A (super-)\emph{Lie $n$-algebra} is a (super-)$L_\infty$-algebra
concentrated in the lowest $n$ degrees.
Given a \mbox{(super-)}$L_\infty$-algebra $\mathfrak{g}$,
we write
$\mathrm{CE}(\mathfrak{g})$ for its \emph{Chevalley-Eilenberg algebra}; which is
a $(\mathbb{Z}\times \mathbb{Z}_2)$-graded commutative dg-algebra
with the property that the underlying graded
super-algebra is the free graded commutative super-algebra on the dual graded super vector space
$\mathfrak{g}[1]^\ast$. These are the dg-algebras which in parts of the
supergravity literature are referred to as ``FDA''s,
a term introduced in \cite{Nieuwenhuizen} and
then picked up in \cite{AFR, DF, CDF} and followups.
Precisely all the (super-)dg-algebras of this \emph{semi-free} form arise as
Chevalley-Eilenberg algebras of (super-)$L_\infty$-algebras this way,
and a homomorphism of $L_\infty$-algebras $f : \mathfrak{g} \to \mathfrak{h}$
is equivalently a homomorphism of dg-algebras of the form
$f^\ast : \mathrm{CE}(\mathfrak{h}) \to \mathrm{CE}(\mathfrak{g})$.
See \cite{frsII} for a review in the context of the higher prequantum geometry
of relevance here and for further pointers to the literature on
$L_\infty$-algebras and their homotopy theory.

\begin{definition}
  For $\mathfrak{g}$ a Lie $n$-algebra, and $X$ a smooth manifold,
  a \emph{flat $\mathfrak{g}$-valued differential form} on $X$
  (of total degree 1, with $\mathfrak{g}$ regarded as cohomologically graded) is equivalently
  a morphism of dg-algebras $A^\ast : \mathrm{CE}(\mathfrak{g}) \to \Omega^\bullet_{\mathrm{dR}}(X)$
  to the de Rham complex.
  Dually we write this as\footnote{The reader familiar with
  $L_\infty$-algebr\emph{oids} should take this as shorthand for the
  $L_\infty$-algebroid homomorphism from the
  tangent Lie algebroid of $X$ to the delooping of the $L_\infty$-algebra $\frak{g}$.}
  $A : X \to \mathfrak{g}$.
  These differential forms naturally pull back along maps of smooth manifolds, and we write
$
  \Omega_{\mathrm{flat}}^1(-,\mathfrak{g})
$
for the sheaf, on smooth manifolds, of flat $\mathfrak{g}$-valued differential forms
of total degree 1.
\label{gValuedDifferentialForms}
\end{definition}
Notice that, in general, these forms of total degree 1 involve differential forms
of higher degree with coefficients in higher degree elements of the $L_\infty$-algebra:
\begin{example}
  For $n \in \mathbb{N}$ write
  $\R[n]$ for the abelian Lie $n$-algebra concentrated
on $\mathbb{R}$ in degree $-n$. Its Chevalley Eilenberg algebra is the dg-algebra
which is genuinely free on a single generator in degree $n+1$. A
flat
$\R[n]$-valued differential form is equivalently just an ordinary
closed differential $(n+1)$-form:
$
  \Omega^1_{\mathrm{flat}}(-, \R[n])
  \simeq
  \Omega^{n+1}_{\mathrm{cl}}
  \,.
$
\end{example}
\begin{definition}
A \emph{$(p+2)$-cocycle} $\mu$ on a Lie $n$-algebra $\mathfrak{g}$ is a degree $p+2$
closed element
in the corresponding Chevalley-Eilenberg algebra $\mu \in \mathrm{CE}(\mathfrak{g})$.
\label{cocycle}
\end{definition}
\begin{remark}
A $(p+2)$-cocycle on $\mathfrak{g}$ is equivalently a map of dg-algebras
$\mathrm{CE}(
\R[p+1])
\to \mathrm{CE}(\mathfrak{g})$ and
hence, equivalently, a map of $L_\infty$-algebras of the form
$\mu : \mathfrak{g} \to \R[p+1]$.
So, if $\{t_a\}$ is a basis for the graded vector space underlying $\mathfrak{g}$,
then the differential $d_{\mathrm{CE}}$ is given in components by
$$
 d_{\mathrm{CE}}\, t^a = \sum_{i \in \mathbb{N}} C^a{}_{a_1 \cdots a_{i}}
  t^{a_1} \wedge \cdots t^{a_{i}}
  \,,
$$
where $\{C^a{}_{a_1 \cdots a_i}\}$ are the structure constants of the $i$-ary
bracket of $\mathfrak{g}$. Consequently, a degree $p+2$ cocycle is a degree $(p+2)$-element
$$
  \mu = \sum_{i} \mu_{a_1 \dots a_i} t^{a_1} \wedge \cdots t^{a_i}
$$
such that
$d_{\mathrm{CE}}\, \mu = 0$.
\end{remark}
\begin{example}
  For $\{t_a\}$ a basis as above and
  $
    \omega \in \Omega^1_{\mathrm{flat}}(X,\mathfrak{g})$
    a $\frak{g}$-valued 1-form on $X$,
  the pullback of the cocycle is the closed differential $(p+2)$-form
  which in  components  reads
  $$
    \mu(\omega)
	  =
	  \sum_{i} \mu_{a_1 \cdots a_i} \omega^{a_1} \wedge \cdots \wedge \omega^{a_i}
	\,,
  $$
  where $\omega^{a} = \omega(t^a)$.
\end{example}
\begin{remark}
Composition $\omega \mapsto (\xymatrix{X \ar[r]^\omega & \mathfrak{g} \ar[r]^-\mu & \mathbb{R}[p+1]})$
of $\mathfrak{g}$-valued differential forms $\omega$ with an $L_\infty$-cocycle
$\mu$ yields a
homomorphism of sheaves
$$
  \Omega^1_{\mathrm{flat}}(-,\mu)
   :
   \xymatrix{
     \Omega_{\mathrm{flat}}(-, \mathfrak{g})
     \ar[r]
	 &
    \Omega^{p+2}_{\mathrm{cl}}
  }
  \,.
$$
This is the sheaf incarnation of $\mu$ regarded as a universal differential form
on the space of all flat $\mathfrak{g}$-valued differential forms. More on this
is below in \ref{Outlook}.
\end{remark}
\begin{example}
By the Yoneda lemma, for $X$ a smooth manifold,
morphisms\footnote{of sheaves, by thinking of $X$ as the sheaf $C^\infty(-, X)$.}
 $X \to \Omega^1_{\mathrm{flat}}(-, \mathfrak{g})$ are equivalently
just flat $\mathfrak{g}$-valued differential forms on $X$.
Specifically, for $G$ an ordinary Lie group, its Maurer-Cartan form
is equivalently a map
$$
  \theta :
  \xymatrix{
    G \ar[r] & \Omega^1_{\mathrm{flat}}(-, \mathfrak{g})
  }
  \,.
$$
Therefore, given a field configuration $\phi : \Sigma_2 \to G$
of the traditional WZW model,
postcomposition with $\theta$ turns this into
$$
  \phi^\ast \theta
    :
  \xymatrix{
    \Sigma
	   \ar[r]^-{\phi}
	&
	G
	\ar[r]^-\theta
	&
	\Omega^1_{\mathrm{flat}}(-,\mathfrak{g})
  }
  \,.
$$
Here if $\mathfrak{g}$ is represented as a matrix Lie algebra then this is the popular
expression $\phi^\ast \theta = \phi^{-1} d \phi$
\label{Yoneda}
\end{example}
\begin{definition}
 Given an $L_\infty$-algebra $\mathfrak{g}$ equipped with a cocycle
 $\mu : \mathfrak{g} \to \mathbb{R}[p+1]$ of degree $p+2$,
 a \emph{perturbative $\sigma$-model datum} for $(\mathfrak{g},\mu)$ is
 a triple consisting of
 \begin{itemize}
   \item a space $X$;
   \item equipped with a flat $\mathfrak{g}$-valued differential form
   $\theta_{\mathrm{global}} : X \to \Omega^1_{\mathrm{flat}}(-,\mathfrak{g})$
    (a ``global Maurer-Cartan form'');
     \item and equipped with a factorization $\mathcal{L}_{\mathrm{WZW}}$
 through $d_{\mathrm{dR}}$ of $\mu(\theta_{\mathrm{global}})$,
 as expressed in the following diagram
$$
  \xymatrix{
	 &
	 X
	 \ar[r]^-{\theta_{\mathrm{global}}}
	 \ar[dr]_{\mathcal{L}_{\mathrm{WZW}}}
	 &
	 \Omega_{\mathrm{flat}}(-, \mathfrak{g})
	 \ar[rr]^-{\mu}
	 &&
	 \Omega^{p+2}_{\mathrm{cl}}\;.
	 \\
	 && \Omega^{p+1}
	 \ar[urr]_{d_{\mathrm{dR}}}
  }
$$
\end{itemize}
The \emph{action functional} associated with this data is the functional
$$
  S_{\mathrm{WZW}}
   :
  \xymatrix{
    [\Sigma,X]
	\ar[r]
	&
	\mathbb{R}
  }
$$
given by
$$
  \phi \mapsto \int_{\Sigma} \mathcal{L}_{\mathrm{WZW}}(\phi)
  \,,
$$
where the integrand is the differential form
$$
  \mathcal{L}_{\mathrm{WZW}}(\phi)
  :
  \xymatrix{
    \Sigma \ar[r]^\phi & X \ar[rr]^-{\mathcal{L}_{\mathrm{WZW}}}
	&&
	\Omega^{p+1}_{\mathrm{cl}}
  }
  \,.
$$
\label{rationaldatum}
\end{definition}

\begin{remark}
Here $X$ actually need not be a (super-)manifold but may be a
smooth higher \mbox{(super-)} stack, hence what we may suggestively
call \emph{higher super orbi-space}.
We make this precise below in section \ref{Outlook}.
\end{remark}
\begin{remark}
 The notation $\theta_{\rm global}$ serves to stress the fact that
we are considering globally defined one-forms on $X$ as opposed to cocycles
in hypercohomology, which is where the higher Maurer-Cartan forms
on \emph{higher} (super-)Lie groups take values, due to presence of
nontrivial higher gauge transformations. See section \ref{Outlook} for more
discussion.
\end{remark}
\begin{remark}
The diagram in Def. \ref{rationaldatum} manifestly captures a local
description, when  $X$ is a contractible manifold.
An immediate global version is captured by the following diagram
$$
  \xymatrix{
     \Sigma \ar[r]^{\eta}
	 &
	 X
	 \ar[r]^-{\theta_{\mathrm{global}}}
	 \ar[dr]_{\mathcal{L}_{\mathrm{WZW}}}
	 &
	 \Omega_{\mathrm{flat}}(-, \mathfrak{g})
	 \ar[rr]^-{\mu}
	 &&
	 \Omega^{p+2}_{\mathrm{cl}}\;,
	 \\
	 && \mathbf{B}^{p+1} U(1)_{\mathrm{conn}}
	 \ar[urr]_{F_{(-)}}
  }
$$
where  $\mathbf{B}^{p+1} U(1)_{\mathrm{conn}}$ is the stack of
$U(1)$-$(p+1)$-bundles with connections,
and $F_{(-)}$ is the curvature morphism; see, for instance, \cite{FSS}.
This globalization is what one sees, for example, in the ordinary
WZW model. This, too, we come to below in section \ref{Outlook}.
\end{remark}

Finally, we notice for discussion in the examples
one aspect of the higher symmetries of such perturbative higher WZW models:
\begin{definition}
  Given a (super-) $L_\infty$-algebra $\mathfrak{g}$,
  its \emph{graded Lie algebra of infinitesimal automorphisms}
  is the Lie algebra whose elements are graded derivations
  $v \in \mathrm{Der}(\mathrm{Sym}^\bullet \mathfrak{g}[1]^\ast)$
  on the graded algebra underlying its Chevalley-Eilenberg algebra
  $\mathrm{CE}(\mathfrak{g})$,
 acting as the corresponding Lie derivatives.
  \label{LieAlgebraOfSymmetries}
\end{definition}

\section{Boundary conditions and brane intersection laws}

In the context of fully extended (i.e. local) topological prequantum field theories,
one has the following general notion of boundary condition, see
\cite{Nuiten, lpqft}.
\begin{definition}
A \emph{prequantum boundary condition for an open brane}
(hence a ``background brane'' on which the given brane may end) is given
by boundary gauge trivializations $\phi_{\mathrm{bdr}}$ of the Lagrangian restricted to
the boundary fields, hence by diagrams of the form
$$
	  \raisebox{23pt}
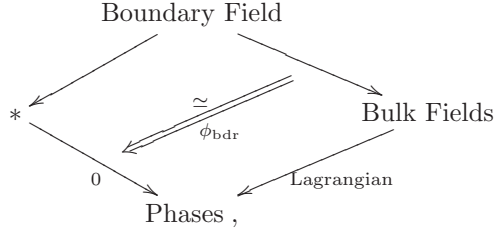
{
	  \xymatrix{
	    & {\rm Boundary~ Field}
		  \ar[dl]
		  \ar[dr]_{\ }="s"
		\\
		\ast
		\ar[dr]_0^{\ }="t"
		&& {\rm Bulk~ Fields} \ar[dl]^{\rm~~ Lagrangian}
		\\
		& {\rm Phases}\;,
		\ar@{=>}^{\phi_{\mathrm{bdr}}}_\simeq "s"; "t"
	  }}
	$$
	where ``Phases" denotes generally the space where the Lagrangian takes values.
	\label{BoundaryConditions}
\end{definition}
Specializing this general principle to our current situation, we have the following
\begin{definition}
  A \emph{boundary condition} for a rational $\sigma$-model datum,
  $(X,\mathfrak{g}, \mu)$ of Def. \ref{rationaldatum}, is
  \begin{enumerate}
    \item an $L_\infty$-algebra $Q$ and a homomorphism
	  $Q \longrightarrow  \mathfrak{g}$,
    \item equipped with a  homotopy $\phi_{\mathrm{brd}}$ of $L_\infty$-algebras
    morphisms
		$
	  \raisebox{23pt}
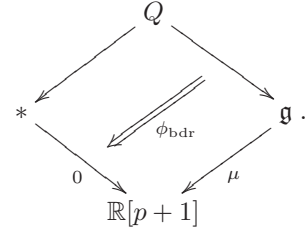
{
	  \xymatrix{
	    & Q
		  \ar[dl]
		  \ar[dr]_{\ }="s"
		\\
		\ast
		\ar[dr]_0^{\ }="t"
		&& \mathfrak{g}\;. \ar[dl]^{\mu}
		\\
		& \mathbb{R}[p+1]
		\ar@{=>}^{\phi_{\mathrm{bdr}}} "s"; "t"
	  }}
	$
  \end{enumerate}
  \label{BoundaryCondition}
\end{definition}
\begin{remark}[Background branes]
 Since $\mathfrak{g}$ is to be thought of as the \emph{spacetime target} for a $\sigma$-model,
 we are to think of $Q \to \mathfrak{g}$ in Def. \ref{BoundaryCondition}
 as a \emph{background brane} ``inside'' spacetime.
 For instance, as demonstrated below in Section \ref{SuperBranesAndTheirIntersectionLaws},
 it may be a D-brane in 10-dimensional super-Minkowski space on which the
 open superstring ends, or it may be the M5-brane in 11-dimensional
 super-Minowski spacetime on which the open M2-brane ends.
 To say then that the $p$-brane described
 by the $\sigma$-model may end on this background brane $Q$
 means to consider worldvolume manifolds $\Sigma_{n}$ with boundaries
 $\partial \Sigma_{p+1} \hookrightarrow \Sigma_{p+1}$
 and \emph{boundary field configurations} $(\phi, \phi|_{\partial})$
 making the left square in the following diagram commute:
$$
  \xymatrix@R=6pt{
    \partial \Sigma_{p+1}
	  \ar[rr]^{\phi|_{\partial \Sigma}}
	\ar[dd]
	&& Q
	 \ar[rr]_{\ }="s" \ar[dd]^{\ }="t"
	&& \ast
	\ar[dd]
    \\
	\\
    \Sigma_{p+1}
	\ar[rr]^-{\phi}
	&&
	\mathfrak{g}
	\ar[rr]_-{\mu}
	&&
	\mathbb{R}[p+1]\;.
	\ar@{=>}^{\mathrm{\phi}_{\mathrm{bdr}}} "s"; "t"
  }
$$
The commutativity of the diagram on the left encodes precisely that
the boundary of the $p$-brane
is to sit inside the background brane $Q$.
But now -- by the defining universal property of the homotopy pullback of
super $L_\infty$-algebras -- this means, equivalently, that the
background brane embedding map
$Q \to \mathfrak{g}$ factors through
the \emph{homotopy fiber}
of the cocycle $\mu$.
If we denote this homotopy fiber by  $\widehat{\mathfrak{g}}$,
then  we have an essentially unique factorization as follows
$$
  \raisebox{20pt}{
  \xymatrix{
    \partial \Sigma_{p+1} \ar[rr]^{\phi|_{\partial \Sigma}} \ar[d]
	&&
	Q  \ar[d]^{\ }
	\ar@{-->}[rr]
	&&
	\widehat{\mathfrak{g}}
	\ar[rr]_{\ }="s"
	\ar[d]^{\ }="t"
	&& \ast
	\ar[d]
    \\
    \Sigma_{p+1}
	\ar[rr]^-{\phi}
	&&
	\mathfrak{g}
	\ar@{=}[rr]
	&&
	\mathfrak{g}
	\ar[rr]_-{\mu}
	&&
	\mathbb{R}[p+1]\;,
	\ar@{=>}^{\phi_{\mathrm{bdr}}^{\mathrm{univ.}}} "s"; "t"
  }
  }
$$
where now $\widehat{\mathfrak{g}} \to \mathfrak{g}$ is the
\emph{homotopy fiber} $\widehat{\mathfrak{g}}$
of the cocycle $\mu$. Notice that here in homotopy theory \emph{all} diagrams
appearing are understood to be filled by homotopies/gauge transformations, but only
if we want to label them explicitly do we display them.
\label{HowToThinkOfExtensionsAsBackgroundBranes}
\end{remark}
The crucial implication to emphasize is that what used to be regarded as a
background brane $Q$ on which the $\sigma$-model brane $\Sigma_n$ may end
is itself characterized by a $\sigma$-model map
$Q \to \widehat{\mathfrak{g}}$, not to the original
target space $\mathfrak{g}$, but to the \emph{extended target space}
$\widehat{\mathfrak{g}}$.  In the class of examples discussed below in
Section \ref{SuperBranesAndTheirIntersectionLaws}, this
extended target space is precisely the
\emph{extended superspace} in the sense of \cite{CdAIP}.

\begin{remark}
  The $L_\infty$-algebra $\widehat{\mathfrak{g}} \to \mathfrak{g}$
  is the \emph{extension} of $\mathfrak{g}$ classified by the cocycle
  $\mu$, in generalization to the traditional extension of Lie algebras
 classified by 2-cocycles.
 If $\mu$ is an $(n_2+1)$-cocycle on an $n_1$-Lie algebra $\mathfrak{g}$
 for $n_1 \leq n_2$, then the extended $L_\infty$-algebra
 $\widehat{\mathfrak{g}}$ is an Lie $n_2$-algebra.
 See \cite{frsII} for more details on this.
 \label{extension}
\end{remark}
\begin{proposition}
The Chevalley-Eilenberg algebra
${\rm CE}(\widehat{\frak{g}})$
of the extension $\widehat{\frak{g}}$ of $\mathfrak{g}$ by a cocycle
$\mu$
 admits, up to equivalence, a very simple description;
namely, it is the differential graded algebra obtained from ${\rm CE}(\frak{g})$
by adding a single generator $c_n$ in degree $n$ subject to the
relation
$$
  d_{{\rm CE}(\widehat{\frak{g}})} \, c_n=\mu
  \,.
$$
Here we are viewing $\mu$ as a degree
$n+1$ element in ${\rm CE}(\frak{g})$, and hence also in
${\rm CE}(\widehat{\frak{g}})$.
\label{CEAlgebrasOfExtensions}
\end{proposition}
\proof
  First observe that we have a commuting diagram of (super-)dg-algebras
  of the form
  $$
    \raisebox{20pt}{
    \xymatrix{
	  \mathrm{CE}\left(\widehat{\mathfrak{g}}\right)
	  &
	  \mathrm{CE}\left(\left(\mathbb{R} \stackrel{\mathrm{id}}{\to} \mathbb{R}\right)[n-1] \right)
	  \ar[l]
	  \\
	  \mathrm{CE}\left(\mathfrak{g}\right)
	  \ar[u]
	  &
	  \mathrm{CE}\left(\mathbb{R}[n]\right)
	  \ar[u]
	  \ar[l]
	}}
	\,.
  $$
  Here the top left dg-algebra is the dg-algebra of the above statement, the
  top morphism is the one that sends the unique degree-$(n+1)$-generator
  to $\mu$ and the unique degree-$n$ generator to $c_n$,
  the vertical morphisms are the evident inclusions, and the
  bottom morphism is the given cocycle. Consider the dual
  diagram of $L_\infty$-algebras
  $$
    \raisebox{20pt}{
    \xymatrix{
	  \widehat{\mathfrak{g}}
	  \ar[r]
	  \ar[d]
	  &
	  (\mathbb{R} \stackrel{\mathrm{id}}{\to} \mathbb{R})[n-1]
	  \ar[d]
	  \\
	  \mathfrak{g}
	  \ar[r]^{\mu}
	  &
	  \mathbb{R}[n]\;.
	}
	}
  $$
  Then observe that the underlying graded vector spaces here
  form a pullback diagram of linear maps
  (the linear components of the $L_\infty$-morphisms).
  From this the statement follows directly with the recognition theorem for
  $L_\infty$-homotopy fibers,
  theorem 3.1.13 in \cite{frsII}.
\endofproof
\begin{remark}
  The construction appearing in  Prop. \ref{CEAlgebrasOfExtensions}
  is of course well familiar in the ``FDA''-technique in the
  supergravity literature \cite{CDF}, and we recall famous examples below in
  Section   \ref{SuperBranesAndTheirIntersectionLaws}. The point to highlight
  here is that this construction has a universal $L_\infty$-homotopy-theoretic meaning,
  in the way described above.
\end{remark}
The crucial consequence of this discussion is the following:
\begin{remark}
If the extension
$\widehat{\mathfrak{g}}$ itself carries a cocycle
$\mu_Q : \widehat{\mathfrak{g}} \to \mathbb{R}[n]$
and  we are able to find a local potential/Lagrangian $\mathcal{L}_{\mathrm{WZW}}$ for the
closed $(n+1)$-form $\mu_Q$ (and we will see in the full description in
\ref{Outlook} that this is always the case),
then this exhibits the background brane $Q$ itself as a rational
WZW $\sigma$-model, now propagating not on the original
``target spacetime'' $\mathfrak{g}$ but on the ``extended spacetime''
$\widehat{\mathfrak{g}}$.
\end{remark}
\begin{remark}
Iterating this process gives rise to a tower of extensions and cocycles
$$
  \raisebox{20pt}{
  \xymatrix{
  \ar@{..}[d] &&
  \\
  \widehat{\widehat{\mathfrak{g}}} \ar[d]
   \ar[rr]^-{\mu_3} && \mathbb{R}[n_3]
  \\
    \widehat{\mathfrak{g}} \ar[d]
     \ar[rr]^-{\mu_2} && \mathbb{R}[n_2]
	\\
	\mathfrak{g} \ar[rr]^-{\mu_1} && \mathbb{R}[n_1]\;,
  }
  }
$$
which is like a
Whitehead tower in rational homotopy theory, only that the cocycles
in each degree here are not required to be the lowest-degree nontrivial ones.
In fact, the actual rational Whitehead tower is an example of this.
In the language of Sullivan's formulation of rational homotopy theory
this says that $\mathfrak{g}_n$ is exhibited by a
sequence of cell attachments as a \emph{relative Sullivan algebra}
relative to $\mathfrak{g}$.
\end{remark}
Since this is an important concept for the present purpose, we give it a name:
\begin{definition}
Given an $L_\infty$-algebra $\mathfrak{g}$,
the {\it brane bouquet of $\frak{g}$} is
the rooted tree consisting of, iteratively, all possible
equivalence classes of nontrivial $\mathbb{R}[\bullet]$ extensions
(corresponding to equivalence classes of nontrivial $\mathbb{R}[\bullet]$-cocycles)
starting with $\mathfrak{g}$ as the root.
\end{definition}
$$
  
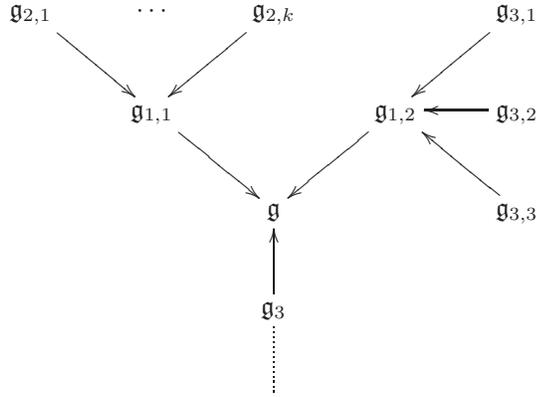
\xymatrix{
     & \mathfrak{g}_{2,1} \ar[dr] & \cdots & \mathfrak{g}_{2,k} \ar[dl]	
	 && \mathfrak{g}_{3,1} \ar[dl]
     \\
     && \mathfrak{g}_{1,1} \ar[dr] && \mathfrak{g}_{1,2} \ar[dl] & \mathfrak{g}_{3,2} \ar[l]
     \\
     &&& \mathfrak{g} & & \mathfrak{g}_{3,3} \ar[ul]
	 \\
	 &&& \mathfrak{g}_3 \ar[u]
     \\
	 &&& \ar@{..}[u]
  }
$$
This \emph{brane bouquet} construction in $L_\infty$-homotopy theory
that we introduced serves to organize and formalize the following two
physical heuristics.
\begin{remark}[Brane intersection laws]
By the discussion above in Remark \ref{HowToThinkOfExtensionsAsBackgroundBranes},
each piece of a brane bouquet of the form
 $$
  \raisebox{20pt}{
  \xymatrix{
    \mathfrak{g}_2 \ar[d]
     \ar[rr]^-{\mu_2} && \mathbb{R}[n_2]
	\\
	\mathfrak{g}_1 \ar[rr]^-{\mu_1} && \mathbb{R}[n_1]
  }
  }
$$
 may be thought of as encoding a
{\it brane intersection law},
 meaning that the WZW $\sigma$-model brane corresponding to
 $(\frak{g}_1, \mu_1)$ can end on the WZW $\sigma$-model brane corresponding
 to $(\frak{g}_2, \mu_2)$.
 Therefore, the brane bouquet of some $L_\infty$-algebra
 $\mathfrak{g}$ lists all the possible $\sigma$-model branes and all
 their intersection laws in the ``target spacetime'' $\mathfrak{g}$.
 \label{BraneIntersectionLaw}
\end{remark}
\begin{remark}[Brane condensates]
 To see how to think of the extensions $\widehat{\mathfrak{g}}$
 as ``extended spacetimes'', observe that by Prop. \ref{CEAlgebrasOfExtensions}
 and Def. \ref{gValuedDifferentialForms} a $\sigma$-model on
 the extension $\widehat{\mathfrak{g}}$ of $\mathfrak{g}$ which is classified by a
 $(p+2)$-cocycle $\mu$ is equivalently a $\sigma$-model on $\mathfrak{g}$
 together with an $p$-form higher gauge field on its worldvolume,
 one whose curvature $(p+1)$-form satisfies a twisted Bianchi identity controled by $\mu$.
 The examples discussed below in Section \ref{SuperBranesAndTheirIntersectionLaws}
 show that this $p$-form field (``tensor field'' in the brane literature)
 is that which is  ``sourced'' by the charged boundaries of the original $\sigma$-model
 branes on $\mathfrak{g}$.
 For instance for superstrings ending on D-branes it is the Chan-Paton gauge field
 sourced by the endpoints of the open string, and for M2-branes ending on
 M5-branes it is the latter's B-field which is sourced by the self-dual strings
 at the boundary of the M2-brane.
 In conclusion, this means that we may think of the extension
 $\widehat{\mathfrak{g}}$ as being the original spacetime $\mathfrak{g}$
 but \emph{filled with a condensate} of branes whose $\sigma$-model
 is induced by $\mu$.
 \label{BraneCondensates}
\end{remark}

\section{Example: Super $p$-branes and their intersection laws}
 \label{SuperBranesAndTheirIntersectionLaws}

 We now discuss higher rational/perturbative WZW models on super-Minkowski spacetime
 regarded as the super-translation Lie algebra over itself,
 as well as on the \emph{extended superspaces} which arise
 as exceptional super Lie $n$-algebra extensions of the super-translation
 Lie algebra. We show then that by the brane intersection laws
 of Remark \ref{BraneIntersectionLaw} this reproduces precisely the
 super $p$-brane content of string/M-theory including the $p$-branes
 with tensor multiplet fields, notably including the D-branes and the M5-brane.
 The discussion is based on the work initiated in
 \cite{DF} and further developed in articles including \cite{CdAIP}.
 The point here is to show that this ``FDA''-technology is
 naturally and usefully reformulated in terms of super-$L_\infty$-homotopy
 theory, and that this serves to clarify and illuminate various points that
 have not been seen, and are indeed hard to see, via the ``FDA''-perspective.

 \smallskip
 We set up some basic notation concerning the super-translation- and
 the super-Poincar{\'e} super Lie algebras, following \cite{DF}. For more background see
 lecture 3 of \cite{FreedLectures} and appendix B of
 \cite{PolchinskiBook}.

 Write $\mathfrak{o}(d-1,1)$ for the Lie algebra of the Lorentz group
in dimension $d$. If $\{\omega_{a}{}^b\}_{a,b}$ is the canonical basis
of Lie algebra elements, then the Chevalley-Eilenberg algebra
$\mathrm{CE}(\mathfrak{o}(d-1,1))$ is generated from elements
$\{\omega^{a}{}_b\}_{a,b}$ in degree $(1,\mathrm{even})$ with the
differential
given by\footnote{Here and in all of the following a summation over repeated indices is
understood.}
$
  d_{\mathrm{CE}}\, \omega^{a}{}_b := \omega^{a}{}_c\wedge \omega^{c}{}_b
 $.
Next, write $\mathfrak{iso}(d-1,1)$ for the Poincar{\'e} Lie algebra.
Its Chevalley-Eilenberg algebra in turn is generated from the
$\{\omega^{a}{}_b\}$
as before together with further generators $\{e^a\}_a$
in degree $(1,\mathrm{even})$ with
differential given by
$
  d_{\mathrm{CE}}\, e^a := \omega^{a}{}_b \wedge e^b
  $.
Now for $N$ denoting a real spinor representation of $\mathfrak{o}(d-1,1)$,
also called the number of supersymmetries (see for instance part 3 of \cite{FreedLectures}),
write $\{\Gamma^a\}$ for a representation of the Clifford algebra
in this representation and $\{\Psi_{\alpha}\}_\alpha$ for the corresponding
basis elements of the spinor representation.
There is then an essentially unique symmetric $\mathrm{Spin}(d-1,1)$-equivariant bilinear map
from two spinors to a vector, traditionally written in components as
$$
  (\psi_1, \psi_2)^a := \frac{i}{2}\overline{\psi} \Gamma^a \psi
  \,.
$$
This induces the
super Poincar{\'e} Lie algebra $\mathfrak{siso}_N(d-1,1)$ whose
Chevalley-Eilenberg super-dg-algebra is generated from the generators as above
together with generators $\{\Psi^{\alpha}\}$ in degree $(1,\mathrm{odd})$
with the differential now defined as follows
  \begin{eqnarray*}
    d_{\mathrm{CE}} \, \omega^{a}{}_b &=& \omega^{a}{}_c \wedge \omega^{c}{}_b\;,
	\\
	d_{\mathrm{CE}} \, e^a &
	  =&
	  \omega^{a}{}_b \wedge e^b + \frac{i}{2}\overline{\psi} \wedge \Gamma^a \psi\;,
	\\
	d_{\mathrm{CE}} \, \psi^\alpha & =& \frac{1}{4} \omega^{a}{}_b \wedge \Gamma^{a}{}_b \psi\;.
  \end{eqnarray*}
Here and in the following $\Gamma^{a_1 \cdots a_p}$ denotes the skew-symmetrized
product of the Clifford matrices and in the above matrix multiplication is
understood whenever the corresponding indices are not displayed.
In summary, the degrees are
$$
  \mathrm{deg}(e^a) = (1, \mathrm{even}),
  \;\;\;\;\;
  \mathrm{deg}(\omega^a) = (1, \mathrm{even}),
  \;\;\;\;\;
  $$
  $$
  \mathrm{deg}(\psi^\alpha) = (1, \mathrm{odd}),
  \;\;\;\;\;
  \mathrm{deg}(d_{\mathrm{CE}}) = (1, \mathrm{even})\;.
$$
Notice that this means that\footnote{
  These grading and sign conventions follow the``Sign manifesto'' in \cite{DeligneFreed}.
  There is another grading and sign convention used in some of the literature, e.g. \cite{BrandtII}.
  Both conventions lead to equivalent cohomology classes.
}
$e^{a_1} \wedge e^{a_2} = - e^{a_1}\wedge e^{a_2}$
and $e^a \wedge \psi^\alpha = - \psi^\alpha \wedge e^a$ but
$\psi^{\alpha_1} \wedge \psi^{\alpha_2} = + \psi^{\alpha_2} \wedge \psi^{\alpha_1}$.
\begin{example}
  For $\Sigma$ a supermanifold of dimension $(d;N)$, a flat $\mathfrak{siso}(d-1,1)$-valued
  differential form $A : \mathrm{CE}(\mathfrak{siso}(d-1,1) \to \Omega^{\bullet}_{\mathrm{dR}}(\Sigma)$, according to
  Def. \ref{gValuedDifferentialForms} and subject to the constraint that the $\mathbb{R}^{d;N}$-component is induced
  from the tangent space of $\Sigma$ (this makes it a \emph{Cartan connection})
  is
  \begin{enumerate}
    \item a \emph{vielbein}  field $E^a := A(e^a)$,
	\item with a \emph{Levi-Civita connection} $\Omega^{a}{}_b := A(\omega^a{}_b)$ (graviton),
	\item a spinor-valued 1-form field $\psi^\alpha := A(\psi^\alpha)$ (gravitino),
  \end{enumerate}
  subject to the flatness constraints which here say that the torsion of
  of the Levi-Civita connection is the super-torsion
  $\tau = \overline{\Psi}\wedge \Gamma^a \Psi \wedge E_a$ and that the
  Riemann curvature vanishes. This is the gravitational field content
  (for vanishing field strength here, one can of course also consider non-flat
  fields; see \cite{SSS})
  of supergravity on $\Sigma$, formulated in
  first order formalism.
  By passing to $L_\infty$-extensions of
  $\mathfrak{siso}$ this is the fomulation
  of supergravity fields which seamlessly generalizes to the higher gauge
  fields that higher supergravities contain, including their correct
  higher gauge transformations. This is the perspective on supergravity
  originating around the article \cite{DF} and expanded on in the textbook
  \cite{CDF}. Recognizing the ``FDA''-language used in this book
  as secretly being about Lie $n$-algebra homotopy theory
  (the ``FDA''s are really Chevalley-Eilenberg algebras super-$L_\infty$-algebras)
  allows to uncover some natural and
  powerful higher gauge theory and geometric homotopy theory \cite{cohesive}
  hidden in traditional supergravity literature.
\end{example}
The \emph{super translation Lie algebra} corresponding to the above is the quotient
$$
  \mathbb{R}^{d;N} := \mathfrak{siso}(d-1,1)/\mathfrak{o}(d-1,1)
$$
whose CE-algebra is as above but with the $\{\omega^{a}{}_b\}$ discarded.
We may think of the underlying super vector space of $\mathbb{R}^{d;N}$ as
$N$-super Minkowski spacetime of dimension $d$, i.e. with $N$ supersymmetries.
Regarded as a supermanifold,
it has canonical super-coordinates $\{x^a, \vartheta^\alpha\}$
and the CE-generators $e^a$ and $\psi^\alpha$ above may be identified
under the general equivalence
$\mathrm{CE}(\mathfrak{g}) \simeq \Omega^\bullet_{\mathrm{L}}(G)$
(for a (super-)Lie group $G$ with (super-)Lie algebra $\mathfrak{g}$)
with the corresponding canonical left-invariant differential forms on
this supermanifold:
 \begin{eqnarray*}
    e^a & = & d_{\mathrm{dR}}\, x^a
	  +
	  \overline{\vartheta} \Gamma^a \,d_{\mathrm{dR}}\, \vartheta\;,
	\\
	\psi^\alpha & = & d_{\mathrm{dR}}\, \vartheta^\alpha\;.
  \end{eqnarray*}
This defines a morphism
$\theta : {\rm CE}(\R^{d;N}) \to \Omega^{\bullet | \bullet}(\R^{d;N})$
to super-differential forms on super Minkowski space,
and via Def. \ref{gValuedDifferentialForms} this is the Maurer-Cartan form,
Example \ref{Yoneda},
on the super\emph{group} $\mathbb{R}^{d;N}$ of supertranslations.
As such $\{e^a, \psi^\alpha\}$ is the canonical \emph{super-vielbein}
on super-Minkowski spacetime.

\smallskip
Notice that the only non-trivial piece of the above CE-differential
remaining on $\mathrm{CE}(\mathbb{R}^{d;N})$ is
$$
  d_{\mathrm{CE}(\mathbb{R}^{d;N})}\, e^a = \overline{\psi} \wedge \Gamma^a\psi
  \,.
$$
Dually this is the single non-trivial super-Lie bracket on $\mathbb{R}^{d;N}$,
the one which pairs two spinors to a vector.
All the exceptional cocycles considered in the following exclusively
are controled by just this equation and Lorentz invariance.

\smallskip

We next consider various branches of the brane bouquet,
Def. \ref{BraneBouquet}, of these super-spacetimes $\mathbb{R}^{d,N}$.

\subsection{$N =1$ $\sigma$-model super $p$-branes --- The old brane scan}
\label{OldBraneScan}

As usual, we write $N$ for a choice of number of irreducible
real (Majorana) representations of $\mathrm{Spin}(d-1,1)$, and
$N = (N_+, N_-)$ if there are two inequivalent chiral minimal representations.
For instance, two important cases are
\begin{center}
\begin{tabular}{|c||c|}
  \hline
  $d = 10$ & $d = 11$
  \\
  \hline
  $N = (1,0) = \mathbf{16}$ & $N = 1 = \mathbf{32}$
  \\
  \hline
\end{tabular}
\end{center}
For $0 \leq p \leq 9$ consider the dual bispinor element
$$
  \mu_p
    :=
  e^{a_1} \wedge \cdots \wedge e^{a_p}
  \wedge
  (\overline{\psi}
  \wedge \Gamma^{a_1 \cdots a_p} \psi)
  \;\in \mathrm{CE}(\mathbb{R}^{d;N})
  \,,
$$
where here and in the following the parentheses are just to guide the reader's eye.
Observe that the differential of this element is of the form
$$
  d_{\mathrm{CE}}\, \mu_p
  \;\propto\;
  e^{a_1} \wedge \cdots \wedge e^{a_{p-1}}
  \wedge
  (
  \overline{\psi}
  \Gamma^{a_1 \cdots a_p}\wedge \psi)
  \wedge
  (\overline{\psi} \wedge \Gamma^{a_p} \psi)
  \,.
$$
This is zero precisely if after skew-symmetrization of the indices, the
spinorial expression
$$
  \overline{\psi}
  \Gamma^{[a_1 \cdots a_p}\wedge \psi
  \wedge
  \overline{\psi} \wedge \Gamma^{a_p]} \psi
  = 0
$$
vanishes identically (on all spinor components).
The spinorial relations which control this are the \emph{Fierz identities}.
If this expression vanishes, then $\mu_p$ is a $(p+2)$-cocycle on
$\mathbb{R}^{d;N=1}$, Def. \ref{cocycle}, hence a homomorphism of
super Lie $n$-algebras of the form
$$
  \mu_p
    :
  \xymatrix{
    \mathbb{R}^{d;N=1}
	\ar[rr]
	&&
	\mathbb{R}[p+1]
  }
  \,.
$$
If this is the case then, by Def. \ref{rationaldatum}, this defines a $\sigma$-model
$p$-brane propagating on $\mathbb{R}^{d;N=1}$.

\smallskip
The combinations of $d$ and $p$ for which this is the case
had originally been worked out in \cite{AETW}.
The interpretation
in terms of super-Lie algebra cohomology was clearly laid out in
\cite{AT}. See \cite{BrandtII, BrandtIII, Brandt} for a rigorous treatment and comprehensive classification for all $N$. The non-trivial
cases (those where $\mu_p$ is closed but not itself a differential) correspond
precisely to the non-empty entries in the following table.\footnote{
The entries show numbers $N$ of of Poincar{\'e} supersymmetries and hence of
minimal real spin representations (Majorana spinors), denoted
$(N,0)$ if they are chiral (Weyl spinors). Notice that in dimensions 5,6 and 7 these
minimal real representations are
``symplectic Majorana'' (hence ``symplectic Majorana-Weyl'' in $d= 6$)  consisting of
two complex irreps, whence they are often instead counted as 2 and $(2,0)$ in the literature.
}\\

\smallskip

\begin{center}
\scalebox{0.7}{ \begin{tabular}{|r||c|c|c|c|c|c|c|c|c|c|}
  \hline
     ${d}\backslash p$ &   & $1$ & $2$ & $3$ & $4$ & $5$ & $6$ & $7$ & $8$ & $9$
	 \\[5pt]
	 \hline \hline
	 $11$ & & &
	  \begin{tabular}{c} (1) \\  $\mathfrak{m}2\mathfrak{brane}$ \end{tabular}  &
	 \hspace{30pt} & \hspace{30pt} &
	 &&& &
	 \\[5pt]
	 \hline
	 $10$ &		
	   & \hspace{-.4cm}\begin{tabular}{c}
	       (1,0) \\  $\mathfrak{string}_{\mathrm{het}}$
		\end{tabular}\hspace{-.4cm}
		&
		&
		&
		&
	  \hspace{-.4cm}\begin{tabular}{c}
	    (1,0) \\ $\mathfrak{ns}5\mathfrak{brane}_{\mathrm{het}}$
	  \end{tabular}\hspace{-.4cm}
	  &
	  &
	  &
	  &
	 \\[5pt]
	 \hline
	 $9$ & & & & & (1) & & & & &
	 \\[5pt]
	 \hline
	 $8$  & & & & (1) & & & & & &
	 \\[5pt]
	 \hline
	 $7$  & & & (1) & & & & & & &
	 \\[5pt]
	 \hline
	 $6$  & & \hspace{-.4cm}\begin{tabular}{c}
	           (1,0) \\  $\mathfrak{littlestring}$
		 \end{tabular} \hspace{-.4cm}
		 & & (1,0) & & & & & &
	 \\[5pt]
	 \hline
	 $5$ & &  & (1) & & & & & & &
	 \\[5pt]
	 \hline
	 $4$  & & (1) & (1) &&&&& & &
	 \\[5pt]
	 \hline
	 $3$  & & (1) &&&&& & & &
	 \\[5pt]
	 \hline
  \end{tabular}
  }
\end{center}

  \vspace{4mm}
This table is known as the ``old brane scan'' for string/M-theory.
Each non-empty entry corresponds to a $p$-brane WZW-type $\sigma$-model action
functional of Green-Schwarz type.
For $(d= 10, p = 1)$ this is the original
Green-Schwarz action functional for the superstring \cite{GreenSchwarz}
and, therefore, we write $\mathfrak{string}_{\mathrm{het}}$ in the respective
entry of the table
(similarly there are cocycles for type II strings, discussed in the following sections),
which at the same time is to denote the
super Lie 2-algebra extension of $\mathbb{R}^{10, N=1}$ that is
classified by $\mu_p$ in this dimension, according to Remark \ref{extension}:
$$
  \xymatrix{
    \mathfrak{string}_{\mathrm{het}}
	\ar[d]
	\\
	\mathbb{R}^{10;N=(1,0)}
	\ar[rr]^-{\mu_1}
	&&
	\mathbb{R}[2]\;.
  }
$$
This Lie 2-algebra has been highlighted in \cite{JH, BHII}.

\smallskip
Analogously we write $\mathfrak{m}2\mathfrak{brane}$ for the
super Lie 3-algebra extension of $\mathbb{R}^{11;N=1}$ classified by
the nontrivial cocycle $\mu_{2}$ in dimension 11
(this was called the \emph{supergravity Lie 3-algebra}
$\mathfrak{sugra}_{11}$ in \cite{SSS})
$$
  \raisebox{20pt}{
  \xymatrix{
    \mathfrak{m}2\mathfrak{brane}
	\ar[d]
	\\
	\mathbb{R}^{11;N=1}
	\ar[rr]^-{\mu_2}
	&&
	\mathbb{R}[3]\;,
  }}
$$
and so on.

\smallskip
While it was a pleasant insight back then that so many of the extended
objects of string/M-theory do appear from just super-Lie algebra cohomology
this way in the above table, it was perhaps just as curious that not all of them
appeared. Later other tabulations of string/M-branes were compiled, based
on less mathematically well defined physical principles \cite{Duff}.
These ``new brane scans'' are what make the above an ``old brane scan''.
But we will show next that if only we allow ourselves to pass from
(super-)Lie algebra theory to (super-) Lie $n$-algebra theory, then
the old brane scan turns out to be part of a brane bouquet that
accurately incorporates all the information of the ``new brane scan'',
all the branes of the new brane scan, altogether with their
intersection laws, with their tensor multiplet field content and its
correct higher gauge transformation laws.

\subsection{Type IIA superstring ending on D-branes and the D0-brane condensate}

We consider the branes in type IIA string theory and point out how
their $L_\infty$-homotopy theoretic formulation serves to
provide a formal statement and proof of the folklore relation between
type IIA string theory with a D0-brane condensate and M-theory.

\smallskip

Write $N = (1,1) = \mathbf{16} + \mathbf{16}^\prime$
for the Dirac representation of $\mathrm{Spin}(9,1)$
given by two 16-dimensional real irreducible representations of opposite chirality.
We write $\{\Gamma^a\}_{a = 1, \cdots, 10}$ for the corresponding
representation of the Clifford algebra and
$\Gamma^{11} := \Gamma^1 \Gamma^2 \cdots \Gamma^{10}$ for the chirality operator.
Finally write $\mathbb{R}^{10;N=(1,1)}$ for the corresponding
super-translation Lie algebra, the super-Minkowski spacetime of type IIA
string theory.

\begin{definition}
The type IIA 3-cocycle is
$$
  \mu_{\mathfrak{string}_{\mathrm{IIA}}}
  :=
  \overline{\psi} \wedge \Gamma^a \Gamma^{11} \psi \wedge e^a
  \;\;:\;\;
  \xymatrix{
    \mathbb{R}^{10;N=(1,1)}
	\ar[r]
	&
	\mathbb{R}[2]
  }
  \,.
$$
The type IIA superstring super Lie 2-algebra is the corresponding
super $L_\infty$-extension
$$
  \xymatrix{
    \mathfrak{string}_{\mathrm{IIA}}
	\ar[d]
	\\
	\mathbb{R}^{10; N=(1,1)}
	\ar[rr]^-{\mu_{\mathfrak{string}_{\mathrm{IIA}}}}
	&&
	\mathbb{R}[2]\;.
  }
$$
Its Chevalley-Eilenberg algebra is that of $\mathbb{R}^{10;N=(1,1)}$
with one generator $F$ in degree $(2,\mathrm{even})$ adjoined and with
its differential being
$$
  d_{\mathrm{CE}}\, F = \mu_{\mathfrak{string}_{\mathrm{IIA}}} =
  \overline{\psi} \wedge \Gamma^a \Gamma^{11} \psi \wedge e^a.
$$
\label{TypeIIALie2Algebra}
\end{definition}

This dg-algebra appears as equation (6.3) in \cite{CdAIP}. It can also be deduced from
 op.cit. that the IIA string Lie 2-algebra of Def. \ref{TypeIIALie2Algebra}
carries exceptional cocycles of
degrees $p+2 \in \{2,4,6,8,10\}$ of the form
  \begin{eqnarray} \label{DBraneCocyclesInIIA}
  \mu_{\mathfrak{d}p\mathfrak{brane}}
    & :=&
	C \wedge e^F
	\\
	& :=&
    \sum_{k=0}^{(p+2)/2} c^p_k
    \left(
      e^{a_1}\wedge \cdots \wedge e^{a_{p-2k}}
    \right)
    \wedge
    \left(
      \overline{\psi}\Gamma^{a_1} \cdots \Gamma^{a_{p-2k}}\Gamma^{11}\psi
    \right)
    \underbrace{F \wedge \cdots \wedge F}_{\mbox{$k$ factors}}\;,
  \end{eqnarray}

where $\{c_k^p \in \mathbb{R}\}$ are some coefficients, and where $C$
denotes the inhomogeneous element of $\mathrm{CE}(\mathbb{R}^{10;N=(1,1)})$
defined by the second line.
For each $p \in \{0,2,4,6,8\}$ there is, up to a global rescaling, a unique choice of
the coefficients $c^p_k$ that make this a cocycle.
This is shown on p. 19 of \cite{CdAIP}.
\begin{remark}
  Here the identification with physics terminology is as follows
  \begin{itemize}
    \item $F$ is the field strength of the \emph{Chan-Paton gauge field}
	on the D-brane, a ``tensor field'' that happens to be a ``vector field'';
	\item $C = \sum_{p} k^p \overline{\psi} \underbrace{e \wedge \cdots \wedge e}_{\mbox{$p$ factors}} \psi$  is the \emph{RR-field}.
  \end{itemize}
\end{remark}
It is interesting to notice the special nature of the cocoycle for the D0-brane:
\begin{remark}
  According to (\ref{DBraneCocyclesInIIA}) for $p = 0$, the cocycle
  defining the D0-brane as a higher WZW $\sigma$-model is just
  $$
    \mu_{\mathfrak{d}0\mathfrak{brane}}
	=
	\overline{\psi}\Gamma^{11}\psi
	\;.
  $$
  Since this independent of the generator $F$, it restricts
  to a cocycle on just $\mathbb{R}^{10;N=(1,1)}$ itself.
  \label{D0cocycleRestrictsToUnextendedSpacetime}
\end{remark}
Concerning this, we highlight the following fact,
which is mathematically elementary but physically noteworthy
(see also Section 2.1 of \cite{CdAIP}),
as it has conceptual consequences for arriving at M-theory starting from
type IIA string theory.
\begin{proposition}
  The extension of 10-dimensional type IIA super-Minkowski spacetime
  $\mathbb{R}^{10;N=(1,1)}$ by the
  D0-brane cocycle as in Remark \ref{D0cocycleRestrictsToUnextendedSpacetime}
  is the 11-dimensional super-Minkowski spacetime of
  11-dimensional supergravity/M-theory:
  $$
    \xymatrix{
	  \mathbb{R}^{11;N=1}
	  \ar[d]
	  \\
	  \mathbb{R}^{10;N=(1,1)}
	  \ar[rr]^-{\mu_{\mathfrak{d}0\mathfrak{brane}}}
	  &&
	  \mathbb{R}[1]\;.
	}
  $$
  \label{11dIsExtensionOf10DByD0Cocycle}
\end{proposition}
\proof
  By Prop. \ref{CEAlgebrasOfExtensions} the Chevalley-Eilenberg algebra of
  the extension classified by
  $\mu_{\mathfrak{d}0\mathfrak{brane}}$ is
  that of $\mathbb{R}^{10;N=(1,1)}$ with one new generator
  $e^{11}$ in degree $(1,\mathrm{even})$ adjoined and with
  its differential defined to be
  $$
    d_{\mathrm{CE}}\, e^{11} = \mu_{\mathfrak{d}0\mathfrak{brane}}
	= \overline{\psi}\Gamma^{11}\psi
	\,.
  $$
  An elementary basic fact of Spin representation theory
  says that the $N = 1$-representation of
  the Spin group $\mathrm{Spin}(10,1)$ in odd dimensions
  is the $N= (1,1)$-representation of
  the even dimensional Spin group $\mathrm{Spin}(9,1)$
  regarded as a representation of the Clifford algebra
  $\{\Gamma^a\}_{a = 1}^{10}$ with $\Gamma^{11}$ adjoined as in
  Def. \ref{TypeIIALie2Algebra}. Using this, the above
  extended CE-algebra is exactly that of $\mathbb{R}^{11;N=1}$
  \,.
\endofproof
\begin{remark}
  In view of Remark \ref{BraneCondensates}
  the content of Prop. \ref{11dIsExtensionOf10DByD0Cocycle}
  translates to heuristic physics language as:
  \emph{A condensate of D0-branes turns the 10-dimensional
  type IIA super-spacetime into the 11-dimensional spacetime of
  11d-supergravity/M-theory.} Alternatively:
\emph{The condensation of D0-branes makes an 11th dimension of
 spacetime appear}.

\smallskip
 In this form the statement is along the lines of the
standard folklore relation between type IIA string theory and M-theory,
which says that type IIA with $N$ D0-branes in it is M-theory
compactified on a circle whose radius scales with $N$;
see for instance \cite{BFSS, Polchinski}. See also
\cite{Kong} for similar remarks motivated from phenomena in 2-dimensional
boundary conformal field theory.
Here in the formalization via higher WZW $\sigma$-models
a version of this statement
becomes a theorem, Prop. \ref{11dIsExtensionOf10DByD0Cocycle}.
  \label{11thDimensionAppearsFromD0Condensate}
\end{remark}
\begin{remark}
  The mechanism of remark \ref{11thDimensionAppearsFromD0Condensate}
  appears at several places in the brane bouquet.
  First of all, since by Prop. \ref{DBraneCocyclesInIIA} the
  D0-brane cocycle is a summand in each type IIA D-brane cocycle,
  it follows via the above translation from $L_\infty$-homotopy theory
  to physics language that:
  \emph{Any type IIA D-brane condensate extends 10-dimensional type IIA
  super-spacetime to 11-dimensional super-spacetime.}
  If we lift attention again from the special case of D-branes of type IIA string theory
  to general higher WZW-type $\sigma$-models, then this mechanism
  is seen to generalize: the 10-dimensional super-Minkowski spacetime
  itself is an extension of the \emph{super-point} by
  10-cocycles (one for each dimension):
  $$
    \xymatrix{
	  \mathbb{R}^{10;N=(1,1)}
	  \ar[d]
	  \\
	  \mathbb{R}^{0;N=(1,1)}
	  \ar[rrr]^{\oplus_{a = 1}^{10}\overline{(-)}\Gamma^a (-)}
	  &&&
	  \mathbb{R}^{10}[1]\;.
	}
  $$
  Here the cocycle describes 10 different 0-brane $\sigma$-models,
  each propagating on the super-point as their target super-spacetime.
  Again, by remark \ref{BraneCondensates}, this mathematical
  fact is a formalization and proof of what in physics language is the
  statement that  \emph{Spacetime itself emerges from the abstract dynamics of 0-branes.}
  This is close to another famous folklore statement about
  string theory. In our context it is a theorem.
\end{remark}

\subsection{Type IIB superstring ending on D-branes and S-duality}

We consider the branes in type IIB string theory
as examples of higher WZW-type $\sigma$-model field theories
and observe how
their $L_\infty$-homotopy theoretic formulation serves to
provide a formal statement of the prequantum S-duality
equivalence between F-strings and D-strings and their unification
as $(p,q)$-string bound states.

\smallskip
Write $N = (2,0) = \mathbf{16} + \mathbf{16}$
for the direct sum representation of $\mathrm{Spin}(9,1)$
given by two 16-dimensional real irreducible representations of
the same chirality.
We write $\{\Gamma^a\}_{a = 1, \cdots, 10}$ for the corresponding
representation of the Clifford algebra on one copy of $\mathbf{16}$
and $\Gamma^a \otimes \sigma^i$ for the linear maps
on their direct sum representation that act as the $i$th Pauli matrix
on $\mathbb{C}^2$ with components $\Gamma^a$, under the canonical identification
$\mathbf{16}\oplus \mathbf{16} \simeq \mathbf{16}\otimes \mathbb{C}^2$.
Finally write $\mathbb{R}^{10;N=(2,0)}$ for the corresponding
super-translation Lie algebra, the super-Minkowski spacetime
of type IIB string theory.

\smallskip
There is a cocycle
$\mu_{\mathfrak{string}_{\mathrm{IIB}}} \in \mathrm{CE}(\mathbb{R}^{10;N=(2,0)})$
given by
$$
  \mu_{\mathfrak{string}_{\mathrm{IIB}}}
  =
  \overline{\psi}\wedge (\Gamma^a \otimes \sigma^3) \psi \wedge e^a
  \,.
$$
The corresponding WZW $\sigma$-model is the Green-Schwarz formulation of the
fundamental type IIB string. Of course we could use
in this formula any of the $\sigma^i$, but one fixed such choice we are
to call the type IIB string. That the other choices are equivalent is
the statement of \emph{S-duality}, to which we come in a moment.
The corresponding $L_\infty$-algebra extension, hence
by Remark \ref{BraneCondensates} the
IIB spacetime ``with string condensate'' is the homotopy fiber
$$
  \xymatrix{
    \mathfrak{string}_{\mathrm{IIB}}
	\ar[d]
	\\
	\mathbb{R}^{10;N=(2,0)}
	\ar[rr]^-{\mu_{\mathfrak{string}_{\mathrm{IIB}}}}
	&&
	\mathbb{R}[2]\;.
  }
$$
As for type IIA,
its Chevalley-Eilenberg algebra $\mathrm{CE}(\mathfrak{string}_{\mathrm{IIB}})$
is that of $\mathbb{R}^{10;N=(2,0)}$ with one generator $F$ in degree
$(2,\mathrm{even})$ adjoined. The differential of that is now given by
  \begin{eqnarray*}
    d_{\mathrm{CE}}\, F
	& =&
	\mu_{\mathfrak{string}_{\mathrm{IIB}}}
	\\
	& = &\overline{\psi} \wedge (\Gamma^a \otimes \sigma^3) \psi \wedge e_a\;.
  \end{eqnarray*}
Now this Lie 2-algebra itself carries exceptional cocycles
of degree $(p+2)$
for $p \in \{1,3,5,7,9\}$ of the form
  \begin{eqnarray} \label{DBraneCocyclesInIIB}
  \mu_{\mathfrak{d}p\mathfrak{brane}}
    & :=&
	C \wedge e^F
	\\
	& :=&
    \sum_{k=0}^{(p+2)/2+1} c^p_k
    \left(
      e^{a_1}\wedge \cdots \wedge e^{a_{p-2k}}
    \right)
    \wedge
    \left(
      \overline{\psi}\wedge (\Gamma^{a_1} \cdots \Gamma^{a_{p-2k}}\otimes \sigma^{1/2})\psi
    \right)
    \underbrace{F \wedge \cdots \wedge F}_{\mbox{$k$ factors}}\;,
  \end{eqnarray}

where on the right the notation $\sigma^{1/2}$
is to mean that $\sigma^1$ appears in summands with an odd number of
generators ``$e$'', and $\sigma^2$
in the other summands.
The corresponding WZW models are those of the type IIB D-branes.

\begin{remark}
  According to expression (\ref{DBraneCocyclesInIIB}) the cocycle of the D1-brane is of
  the form
  $$
    \mu_{\mathfrak{d}1\mathfrak{}brane}
	=
	\overline{\psi} \wedge (\Gamma^a \otimes \sigma^1) \wedge e^a
	\,,
  $$
  which is the same form as that of $\mu_{\mathfrak{string}_{\mathrm{IIB}}}$
  itself, only that $\sigma^3$ is replaced by $\sigma^1$.
  In fact since this is the D-brane cocycle which is
  independent of the new generator $F$, it restricts to a cocycle
  on just $\mathbb{R}^{10;N=(2,0)}$ itself.
  So the cocycle for the ``F-string'' in type IIB is on the same
  footing as that of the ``D-string''. Both differ only by a
  ``rotation'' in an internal space.
\end{remark}
\begin{remark}
There is a circle worth of $L_\infty$-automorphisms
$$
  S(\alpha) : \mathbb{R}^{10;N=(2,0)}\to \mathbb{R}^{10;N=(2,0)}\;,
$$
hence a group homomorphism
$$
  U(1) \to \mathrm{Aut}(\mathbb{R}^{10;N=(2,0)})\;,
$$
given dually on Chevalley-Eilenberg algebras by
  \begin{eqnarray*}
    e^a &\mapsto&  e^a
	\\
	\psi &\mapsto& \exp(\alpha \sigma^2) \psi\;.
  \end{eqnarray*}
This mixes the cocycles for the F-string and for the D-string in that
for a quarter rotation it turns one into the other
$$
    S(\pi/4)^\ast (\mu_{\mathfrak{string}_{\mathrm{IIA}}})
     = \mu_{\mathfrak{d}1\mathfrak{brane}}\;,
$$
and for a rotation by a general angle it produces a corresponding
superposition of both.
In particular, we can form \emph{bound states} of $F$-strings and D1-branes
by adding these cocycles
$$
  \mu_{(p,q)\mathfrak{string}}
  =
  p \,\mu_{\mathfrak{string}_{\mathrm{IIB}}}
  +
  q \,\mu_{\mathfrak{d}1\mathfrak{brane}}
  \;\;
  \in \mathrm{CE}(\mathbb{R}^{10;N=(2,0)})
  \,.
$$
These define the $(p,q)$-string bound states as WZW-type $\sigma$-models.
\end{remark}

\subsection{The M-theory 5-brane and the M-theory super Lie algebra}
\label{The5brane}

We discuss here the single M5-brane as a higher WZW-type $\sigma$-model,
show that it is defined by a 7-cocycle on the M2-brane super Lie-3 algebra
and observe that this 7-cocycle is indeed the relevant fermionic
7d Chern-Simons term of 11-dimensional supergravity compactified on
$S^4$, as required by $\mathrm{AdS}_7/\mathrm{CFT}_6$
in the Chern-Simons interpretation of \cite{Witten}. We see that the
truncation of the symmetry algebra of this higher 5-brane superalgebra
to degree 0 is the ``M-algebra''.

\smallskip

Write $N = 1 = \mathbf{32}$ for the irreducible real representation of
$\mathrm{Spin}(10,1)$. Write  $\{\Gamma^a\}_{a = 1}^{11}$ for the
corresponding representation of the Clifford algebra. Finally write
$\mathbb{R}^{11; N =1}$ for the corresponding super-translation Lie algebra.
According to the old brane scan in section \ref{OldBraneScan},
the exceptional Lorentz-invariant
cocycle for the M2-brane is

$$
  \mu_{\mathfrak{m}2\mathfrak{brane}}
    =
  \overline{\psi} \wedge \Gamma^{a b} \psi \wedge e^a \wedge e^b\;.
$$
The Green-Schwarz action functional for the M2-brane is the $\sigma$-model defined
by this cocycle
$$
  \xymatrix{
    \mathbb{R}^{11;N}
	\ar[rr]^{\mu_{\mathfrak{m}2\mathfrak{brane}}}
	&&
	\mathbb{R}[3]\;.
  }
$$

By the $L_\infty$-theoretic brane intersection law
of Remark \ref{BraneIntersectionLaw}, for the M2-brane to end on another kind
of brane, that other WZW model is to have
the extended spacetime $\mu_{\mathfrak{m}2\mathfrak{brane}}$
(the original spacetime including a condensate of M2s) as its
target space.
By Prop. \ref{CEAlgebrasOfExtensions},
  the Chevalley-Eilenberg algebra of the M2-brane algebra is
  obtained from that of the super-Poincar{\'e} Lie algebra by
  adding one more generator $c_3$ with $\mathrm{deg}(c_3) = (3,\mathrm{even})$
  with differential defined by
    \begin{eqnarray*}
    d_{\mathrm{CE}} \,c_3
	& :=&
	\mu_{\mathfrak{m}2\mathfrak{brane}}
	\\
	&=&
	\overline{\psi} \wedge \Gamma^{a b} \psi \wedge e^a \wedge e^b\,.
	\end{eqnarray*}
We can then define an extended spacetime Maurer-Cartan form $\hat \theta$ in
$\Omega^1_{\rm flat}(\R^{11;N}, \mathfrak{m2brane})$,
extending the canonical Maurer-Cartan form $\theta$ in
$\Omega^1_{\rm flat}(\R^{11;N}, \R^{d;N})$,
by picking any 3-form $C_3 \in \Omega^3(\mathbb{R}^{11;N})$ such that
$d_{\mathrm{dR}} C_3 = \overline{\psi} \Gamma^{a b} \wedge \psi \wedge e^a \wedge e^b$.

\smallskip
Next, for every $(n+1)$ cocycle on $\mathfrak{m}2\mathfrak{brane}$ we get
an $n$-dimensional WZW model defined on $\mathbb{R}^{11;N}$ this way.
In particular, the next one we meet is the M5-brane cocycle.
  Indeed, there is the degree-7 cocycle
  $$
    \mu_7
	  =
	\overline{\psi} \Gamma^{a_1 \cdots a_5} \psi e^{a_1} \wedge \cdots e^{a_5}
	 +
	C_3 \wedge \overline{\psi} \Gamma^{a b} \psi \wedge e^a \wedge e^b
	\;:\;
	\xymatrix{
	 \mathfrak{m}2\mathfrak{brane}
	 \ar[rr]
	 &&
	 \mathbb{R}[6]
	}
	\,
  $$
that was first observed in \cite{DF}, then rediscovered several times, for instance in
  \cite{Sezgin}, in \cite{BLNPST1} and in \cite{CdAIP}.
  Here we identify it as an $L_\infty$ 7-cocycle on
  the $\mathfrak{m2brane}$ super Lie 3-algebra. The
  $L_\infty$-extension of $\mathfrak{m2brane}$ associated with the 7-cocycle is
  a super Lie 6-algebra that we call  $\mathfrak{m5brane}$.

\smallskip
It follows from this, with remark \ref{BraneIntersectionLaw},
that the M2-brane may end on a M5-brane whose WZW term
$\mathcal{L}_{\mathrm{WZW}}$ locally satisfies
$$
  d
  \mathcal{L}_{\mathrm{WZW}}
  =\mu_7=
  \overline{\psi} \Gamma^{a_1 \cdots a_5} \psi e^{a_1} \wedge \cdots e^{a_5}
	 +
  C_3 \wedge \overline{\psi} \Gamma^{a b} \psi \wedge e^a \wedge e^b
$$
This is precisely what in \cite{BLNPST1} is argued to be the
action functional of the M5-brane (here displayed in the absence of the
bosonic contribution of the C-field).
However, in order to get the expected structure of gauge transformations,
we need to go further. Namely, while the above
local expression for the action functional appears to be correct on the nose,
its  gauge transformations are not as expected for the M5: for the
M5-brane worldvolume theory the 2-form with curvature $C_3$ is
supposed to be a genuine higher 2-form gauge field on the worldvolume,
directly analogous to the Neveu-Schwarz B-field of 10-dimensional
supergravity spacetime; see \cite{7d}.
 As such, it is to have gauge transformations
parameterized by 1-forms. But in the above formulation fields are maps
$\Sigma_6 \to \mathbb{R}^{11;N}$ into spacetime itself, and as such
have no gauge transformations at all.
We can fix this by finding a better space $\hat X$. In fact we should take that
to be $\mathfrak{m}2\mathfrak{brane}$ itself. As indicated above, this is an extension
$$
  \xymatrix{
   \mathbb{R}[2] \ar[r]
	 &
	 \mathfrak{m}2\mathfrak{brane}
	  \ar[r]
	 &
	 \mathbb{R}^{11; N}
  }\;,
$$
and, hence, a twisted product of spacetime with $\mathbb{R}[2]$,
the infinitesimal version of the moduli space of 2-form connections.
 We see this more precisely below in Section \ref{Outlook}.

\begin{remark}
  By $\mathrm{AdS}_7/\mathrm{CFT}_6$ duality and by \cite{Witten} the M5-brane is supposed to be
  the 6-dimensional WZW model which is holographically related to
  the 7-dimensional Chern-Simons term inside 11-dimensional supergravity
  compactified on a 4-sphere in analogy to how the traditional 2d WZW model
  is the holographic dual of ordinary 3d Chern-Simons theory.
  By our discussion here that 7d Chern-Simons theory ought to be the
  one given by the 7-cocycle. Indeed, we observe that this 7-cocycle
  does appear in the compactification according to D'Auria-Fre \cite{DF}.
  Back in that article these authors worked locally and discarded precisely
  this term as a global derivative, but in fact it is a topological term
  as befits a Chern-Simons term and may \emph{not} be discarded globally.
  This connects the discussion here to the holographic
  $\mathrm{AdS}_7$/$\mathrm{CFT}_6$-description of the \emph{single} M5-brane.
  Now a coincident $N$-tuple of M5-branes is supposed to be determined by
  a semisimple Lie algebra and nonabelian higher gauge field data.
  Since $\mathrm{AdS}_7/\mathrm{CFT}_6$ is still supposed to apply, we are
  to consider the \emph{nonabelian} contributions to the 7-dimensional
  Chern-Simons term in 11d sugra compactified to $\mathrm{AdS}_7$.
  These follow from the 11-dimensional anomaly cancellation and charge
  quantization. Putting this together as discussed in \cite{7d,CField}
  yields the corresponding 7d Chern-Simons theory. Among other terms it is controled
  by the canonical 7-cocycle $\mu_{7}^{\mathfrak{so}}$
  on the semisimple Lie algebra $\mathfrak{so}$. Since this extends
  evidently to a cocycle also on the super Poincar{\'e} Lie algebra,
  we may just add it to the bispinorial cocycle that defines the single M5,
  to get
  $$
    \xymatrix{
	  \mathbb{R}^{11;N=1} \times \mathfrak{so}(10,1)
	  \ar[rrrrr]^-{\overline{\psi}e^5\psi
	   +
	   \langle \omega \wedge [\omega \wedge \omega] \wedge [\omega \wedge \omega]\wedge [\omega \wedge \omega]  \rangle}
	  &&&&&
	  \mathbb{R}[6]
	}
	\,.
  $$
  By the general theory indicated here this defines a 6-dimensional
  WZW model. By the discussion in \cite{7d, CField} it satisfies
  all the conditions imposed by holography. It is to be expected that this
  is part of the description of the nonabelian M5-brane.
\end{remark}

Finally it is interesting to consider the symmetries of the M5-brane
higher WZW model obtained this way.
\begin{definition}
  \label{11dPoincarePolyvectorExtension}
  The \emph{polyvector extension} \cite{ACDP} of $\mathfrak{sIso}(10,1)$
  -- called the \emph{M-theory Lie algebra} \cite{Sezgin} --
  is the super Lie algebra obtained by adjoining to $\mathfrak{sIso}(10,1)$ generators
  $\{Q_\alpha, Z^{ab}\}$ that transform as spinors with respect to the
  existing generators, and whose non-vanishing brackets among themselves are
 \begin{eqnarray}
    \left[ Q_\alpha, Q_\beta \right] &=& i(C \Gamma^a)_{\alpha \beta} P_a
	+ (C \Gamma_{a b}) Z^{a b}\;,
	\nonumber\\
    \left[ Q_\alpha, 	Z^{ab}\right] &= &2 i (C \Gamma^{[a})_{\alpha \beta} Q^{b]\beta}
	\,.
	\nonumber
  \end{eqnarray}
\end{definition}
\begin{proposition}
  The degree-0 piece of the
  graded Lie algebra of infinitesimal automorphisms
  of $\mathfrak{m}2\mathfrak{brane}$, Def. \ref{LieAlgebraOfSymmetries},
  is the ``M-theory algebra'' polyvector extension of the 11d super Poincar{\'e} algebra of
  Def. \ref{11dPoincarePolyvectorExtension}.
\end{proposition}
\proof
  We leave this as an exercise to the reader.
  Hint: under the identification of FDA-language with ingredients
  of $L_\infty$-homotopy theory as discussed here, one can see that
  this involves the computations displayed in \cite{Castellani}.
\endofproof

\subsection{The complete brane bouquet of string/M-theory}

We have discussed various higher super Lie $n$-algebras of
super-spacetime. Here we now sum up, list all the relevant extensions
and fit them into the full brane bouquet.
To state the brane bouquet, we first need names for all the branches that
it has

\begin{definition}
The \emph{refined brane scan} is the following collection of
values of triples $(d,p,N)$.
%

\vspace{2mm}
\hspace{0.1cm}
\scalebox{0.65}{
\begin{tabular}{|r||c|c|c|c|c|c|c|c|c|c|ccccc|}
  \hline
  &&&&&&&&&&
  \\
     ${d}\backslash p$ & $0$ & $1$ & $2$ & $3$ & $4$ & $5$ & $6$ & $7$ & $8$ & $9$
	 \\
	 \hline \hline
	 $11$ & & &
	  \hspace{-.4cm}\begin{tabular}{ll} (1)
	    & $\mathfrak{m}2\mathfrak{brane}$ \end{tabular}\hspace{-.4cm} &
	 \hspace{30pt} & \hspace{30pt} &
	 \begin{tabular}{ll} (1) \hspace{-.4cm} & $\mathfrak{m}5\mathfrak{brane}$ \end{tabular} &&&&
	 \\
	 \hline
	 $10$ &		
	   \hspace{-.4cm}
	   \begin{tabular}{cc}
		  (1,1) \\ \hspace{0cm}  $\mathfrak{D}0\mathfrak{brane}$
		\end{tabular}
	   \hspace{-.4cm}
	 & \hspace{-.3cm}\begin{tabular}{ll}
	       (1,0) & \hspace{-.4cm} $\mathfrak{string}_{\mathrm{het}}$
    	\\ (1,1) & \hspace{-.4cm} $\mathfrak{string}_{\mathrm{IIA}}$
     	\\ (2,0) & \hspace{-.4cm} $\mathfrak{string}_{\mathrm{IIB}}$
		\\ (2,0) & \hspace{-.4cm} $\mathfrak{D}1\mathfrak{brane}$
		\end{tabular}\hspace{-.4cm}
		&
		\hspace{-.4cm}\begin{tabular}{cc}
		  (1,1) \\  $\mathfrak{D}2\mathfrak{brane}$
		\end{tabular}\hspace{-.4cm}
		&
		\hspace{-.4cm}\begin{tabular}{cc}
		  (2,0)\\ $\mathfrak{D}3\mathfrak{brane}$
		\end{tabular}\hspace{-.4cm}
		&
		\hspace{-.4cm}\begin{tabular}{cc}
		  (1,1) \\ $\mathfrak{D}4\mathfrak{brane}$
		\end{tabular}\hspace{-.4cm}
		&
	  \hspace{-.4cm}\begin{tabular}{ll}
	    (1,0) & \hspace{-.4cm} $\mathfrak{ns}5\mathfrak{brane}_{\mathrm{het}}$
		\\
	    (1,1) & \hspace{-.4cm} $\mathfrak{ns}5\mathfrak{brane}_{\mathrm{IIA}}$
	    \\
	    (2,0) & \hspace{-.4cm} $\mathfrak{ns}5\mathfrak{brane}_{\mathrm{IIB}}$
		\\
		(2,0) & \hspace{-.4cm} $\mathfrak{D}5\mathfrak{brane}$
	  \end{tabular}\hspace{-.4cm}
	  &
		\hspace{-.4cm}\begin{tabular}{cc}
		  (1,1) \\$\mathfrak{D}6\mathfrak{brane}$
		\end{tabular}\hspace{-.4cm}
	  &
		\hspace{-.4cm}\begin{tabular}{cc}
		  (2,0) \\  $\mathfrak{D}7\mathfrak{brane}$
		\end{tabular}\hspace{-.4cm}
	  &
		\hspace{-.4cm}\begin{tabular}{cc}
		  (1,1) \\ $\mathfrak{D}8\mathfrak{brane}$
		\end{tabular}\hspace{-.4cm}
	  &
		\hspace{-.4cm}\begin{tabular}{cc}
		  (2,0) \\ $\mathfrak{D}9\mathfrak{brane}$
		\end{tabular}\hspace{-.4cm}
	 \\
	 \hline
	 $9$ & & & & & (1) & & & & &
	 \\
	 \hline
	 $8$  & & & & (1) & & & & & &
	 \\
	 \hline
	 $7$  & & & (1) & & & & & & &
	 \\
	 \hline
	 $6$  & & \hspace{-.4cm}\begin{tabular}{ll}
 	           (1,0) & \hspace{-.4cm} $\mathfrak{sdstring}$
		 \end{tabular} \hspace{-.4cm}
		 & & (1,0) & & & & & &
	 \\
	 \hline
	 $5$ & &  & (1) & & & & & &&
	 \\
	 \hline
	 $4$  & & (1) & (1) &&&& && &
	 \\
	 \hline
	 $3$  & & (1) &&& && & &&
	 \\
	 \hline
  \end{tabular}
  }
  \label{refinedbranescan}
\end{definition}


\noindent The entries of this table denote super-$L_\infty$-algebras
that organize themselves
as nodes in the brane bouquet
according to the following proposition.

\begin{proposition}[The brane bouquet]
There exists a system of higher super-Lie-$n$-algebra extensions of the
super-translation Lie algebra $\mathbb{R}^{d;N}$
for $(d = 11, N=1)$,  $(d = 10, N=(1,1))$, for $(d = 10, N = (2,0))$
and for $(d = 6, N = (2,0))$,
which is jointly given by the following diagram
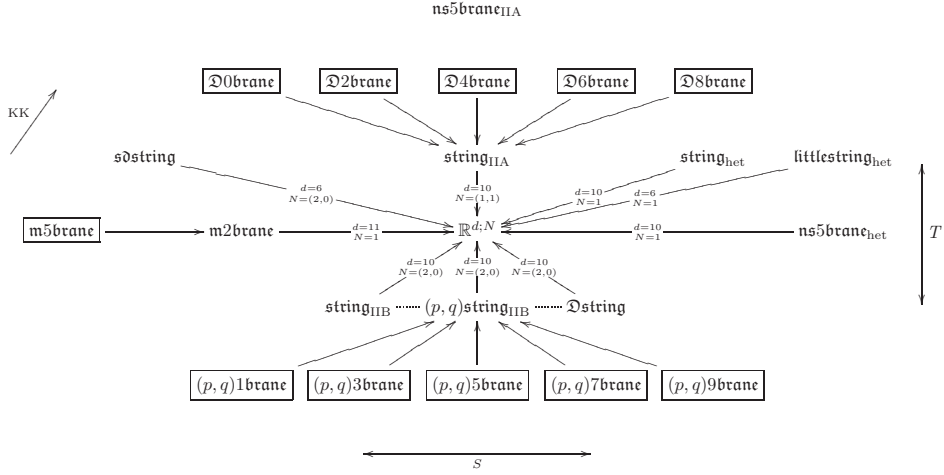
\centerline{\scalebox{0.68}{
$$
  \hspace{-.6cm}
  \xymatrix@C=2pt{
    && && && \mathfrak{ns}5\mathfrak{brane}_{\mathrm{IIA}}
    \\
    &&
	&& \fbox{$\mathfrak{D}0\mathfrak{brane}$} \ar[drr]
	& \fbox{$\mathfrak{D}2\mathfrak{brane}$} \ar[dr]
	& \fbox{$\mathfrak{D}4\mathfrak{brane}$} \ar[d]
	& \fbox{$\mathfrak{D}6\mathfrak{brane}$} \ar[dl]
	& \fbox{$\mathfrak{D}8\mathfrak{brane}$} \ar[dll]
    \\
    & \ar[ur]^{\mathrm{KK}}&
	& \mathfrak{sdstring} \ar[drrr]|{{d = 6} \atop {N = (2,0)}}
	&
	&& \mathfrak{string}_{\mathrm{IIA}} \ar[d]|-{{d=10} \atop {N=(1,1)}}
	&& \mathfrak{string}_{\mathrm{het}} \ar[dll]|-{{d=10}\atop {N = 1}}
	&& \mathfrak{littlestring}_{\mathrm{het}} \ar[dllll]|-{{d=6}\atop {N = 1}}
	&&
     \ar@{<->}[dd]^{\mbox{T}}	
	&&
    \\
    && \fbox{$\mathfrak{m}5\mathfrak{brane}$} \ar[rr]
	&& \mathfrak{m}2\mathfrak{brane} \ar[rr]|-{{d=11} \atop {N=1}}
	&& \mathbb{R}^{d;N}
	&&
	&& \mathfrak{ns}5\mathfrak{brane}_{\mathrm{het}} \ar[llll]|-{{d = 10}\atop {N =1}}
	&&
	\\
	&&
	&&
	& \mathfrak{string}_{\mathrm{IIB}} \ar[ur]|-{{d = 10}\atop {N=(2,0)}}
	\ar@{.}[r]
	& (p,q)\mathfrak{string}_{\mathrm{IIB}} \ar[u]|-{{d = 10}\atop {N=(2,0)}}
	\ar@{.}[r]
	& \mathfrak{Dstring} \ar[ul]|-{{d = 10}\atop {N=(2,0)}}
	&&
	&&
	&&
    \\
    &&
	&& \fbox{$(p,q)1\mathfrak{brane}$} \ar[urr]
	&\fbox{$(p,q)3\mathfrak{brane}$} \ar[ur]
	& \fbox{$(p,q)5\mathfrak{brane}$} \ar[u]
	& \fbox{$(p,q)7\mathfrak{brane}$} \ar[ul]
	& \fbox{$(p,q)9\mathfrak{brane}$} \ar[ull]
	\\
	&& &&  & \ar@{<->}[rr]_S &  &&
  }
$$
}
}
where
\begin{itemize}
  \item
    An object in this diagram is precisely a super-Lie-$(p+1)$-algebra
	extension of the super translation algebra $\R^{d;N}$, with
	$(d,p,N)$ as given by the entries of the same name in the
	refined brane scan, def. \ref{refinedbranescan};
  \item every morphism
  is a super-Lie $(p+1)$-algebra extension by an exceptional $\mathbb{R}$-valued
  $\mathfrak{o}(d)$-invariant super-$L_\infty$-cocycle
  of degree $p+2$ on the domain of the morphism;
  \item
    the unboxed morphisms are hence super Lie $(p+1)$-algebra extensions of
	$\mathbb{R}^{d;N}$ by a super Lie algebra $(p+2)$-cocycle, hence are homotopy
	fibers
	of the form
	$$
	  \xymatrix{
	     p\mathfrak{brane}
		 \ar[d]
		 \ar[rr]_<{\rfloor}
		 &&
		 \ast
		 \ar[d]
		 \\
		 \mathbb{R}^{d;N}
		 \ar[rr]^-{\mathrm{some\;cocycle}}
		 &&
		 \mathbb{R}[p+1]\;,
	  }
	$$
  \item and the boxed super-$L_\infty$-algebras are super Lie $(p+1)$-algebra
  extensions of genuine super-$L_\infty$-algebras (which are not plain super Lie algebras),
  again by $\mathbb{R}$-cocycles
	$$
	  \xymatrix{
	     p_2\mathfrak{brane}
		 \ar[d]
		 \ar[rr]_<{\rfloor}
		 &&
		 \ast
		 \ar[d]
		 \\
		 p_1\mathfrak{brane}
		 \ar[rr]^-{\mathrm{some\;cocycle}}
		 &&
		 \mathbb{R}[p_2+1]\;.
	  }
	$$
\end{itemize}
 \label{BraneBouquet}
\end{proposition}
\proof
  Using prop. \ref{CEAlgebrasOfExtensions}
  and the dictionary that we have established above
  between the language used in the
  physics literature (``FDA''s) and super-$L_\infty$-algebra homotopy theory,
  this is a translation of the following results that
  can be found scattered in the literature (some of which were discussed in the
  previous sections).
  \begin{itemize}
  \item All $N=1$-extensions of $\mathbb{R}^{d;N=1}$
    are those corresponding to the
  ``old brane scan'' \cite{AETW}.
  Specifically the cocycle which classifies the super Lie 3-algebra extension
  $\mathfrak{m}2\mathfrak{brane} \to \R^{11;1}$
  had been found earlier in the context of supergravity around
  equation
  (3.12) of \cite{DF}. These authors also explicitly write down
  the ``FDA'' that then in \cite{SSS} was recognized as the
  Chevalley-Eilenberg algebra of the super Lie 3-algebra
  $\mathfrak{m}2\mathfrak{brane}$ (there called the ``supergravity Lie 3-algebra'').
   Later all these cocycles appear in the
  systematic classification of super Lie algebra cohomology
  in \cite{BrandtII, BrandtIII, Brandt}.

  \item The 7-cocycle classifying the super-Lie-6-algebra
   extension $\mathfrak{m}5\mathfrak{brane} \to \mathfrak{m}2 \mathfrak{brane}$
   together with that extension itself
   can be traced back, in FDA-language,
   to (3.26) in \cite{DF}. This is maybe still the only previous
   reference that makes explicit the Lie 6-algebra extension (as an ``FDA''),
   but the corresponding 7-cocycle itself has later been rediscovered several times,
   more or less explicitly.  For instance it appears as equations (6) and (9) in
   \cite{BLNPST1}.
   A systematic discussion is
   in section 8 of \cite{CdAIP}.

   \item The extension $\mathfrak{string}_{\mathrm{IIA}} \to \mathbb{R}^{10;N=(1,1)}$
   by a super Lie algebra 3-coycle and the cocycles for the further higher extensions
   $\mathfrak{D}(2n)\mathfrak{brane} \to \mathfrak{string}_{\mathrm{IIA}}$
can be traced back to section 6 of \cite{CdAIP}.

   \item The extension $\mathfrak{string}_{\mathrm{IIB}} \to \mathbb{R}^{10;N=(2,0)}$
   by a super Lie algebra 2-coycle and the cocycles for the further higher extensions
   $\mathfrak{D}(2n+1)\mathfrak{brane} \to \mathfrak{string}_{\mathrm{IIA}}$,
   as well as the extension
   $\mathfrak{ns}5\mathfrak{brane}_{\mathrm{IIB}} \to \mathfrak{Dstring}$
   follow from section 2 of \cite{Sak2}.

   \end{itemize}
\endofproof
\begin{remark}
  The look of the brane bouquet, Prop. \ref{BraneBouquet},
  is reminiscent of the famous cartoon that displays the
  conjectured coupling limits of string/M-theory,
  e.g. figure 4 in \cite{cartoon}, or fig. 1 in \cite{Polchinski}.
  Contrary to that cartoon, the brane bouquet is a theorem.
  Of course that cartoon alludes to more details
  of the nature of string/M-theory than we are
  currently discussing here, but all the more should it
  be worthwhile to have a formalism that makes precise
  at least the basic structure, so as to be able to proceed
  from solid foundations.
\end{remark}

\section{Non-perturbative higher WZW models on higher super-orbispaces}
 \label{Outlook}

In this final section
we give a non-perturbative (globalized) refinement of the perturbative higher WZW-models that
we discussed so far. These non-perturbative higher WZW models are naturally formulated
not just in higher Lie theory as used so far, but
in genuine higher differential geometry, which means in higher smooth
and supergeometric \emph{stacks}. In the language of physics, stacks
may best be thought of as \emph{higher orbispaces}, the generalization
of \emph{orbifolds} and more generally of orbispaces (dropping the
finiteness condition) to the case where there
are not just gauge transformations between points, but also higher gauge
transformations between these. The idea of considering $\sigma$-models
on orbifold target spaces is traditionally familiar, and here we
generalize this naturally by allowing these target spaces
to be such higher \mbox{(super-)}orbispaces.
The reader can find an exposition of the technology relevant for the following
in \cite{StackyPerspective}, a collection of all the relevant definitions
and constructions in \cite{hgp}, and the full technical details in \cite{cohesive}.

\smallskip

In higher (super-)differential geometry every (super-) $L_\infty$-algebra
$\mathfrak{g}$ has \emph{Lie integrations} to higher smooth (super-)groups
$G$; see \cite{FSS} for details. (For
$\mathfrak{g} = \mathfrak{string}_{\mathrm{het}}$ the Lie integration is
discussed in \cite{JHIII}.)
For instance, the abelian $L_\infty$-algebra $\mathbb{R}[n]$ integrates
to the \emph{circle n+1-group} $\mathbf{B}^n U(1)$. This is at the same
time the higher \emph{moduli stack} for circle $n$-bundles
(also called $(n-1)$-bundle gerbes).

\smallskip
Recall then from the Introduction that
 a perturbative higher WZW model of dimension $n$ is all encoded by a
morphism of \mbox{(super-)}$L_\infty$-algebras of the form
$$
  \mu \;:\; \xymatrix{ \mathfrak{g} \ar[r] &  \mathbb{R}[n]}
  \,.
$$
Therefore, its non-perturbative refinement is to be an $n$-form connection on
a circle $n$-bundle over the higher group $G$. The latter is given by a morphism of
higher smooth \mbox{(super-)}groups the form
$$
  \Omega \mathbf{c}
  :
  \xymatrix{
    G \ar[r] & \mathbf{B}^n U(1)
  }
  \,.
$$
(This is the higher and smooth analog of the canonical morphism $G\to K(\mathbb{Z},3)$ defining the fundamental class $[\omega_G]\in H^3(G;\mathbb{Z})$ for a compact, simple
and simply connected Lie group $G$, in the traditional WZW model.)
Equivalently, this is a morphism of the corresponding delooping stacks
$$
  \mathbf{c}
  :
  \xymatrix{
    \mathbf{B}G \ar[r] & \mathbf{B}^{n+1} U(1)
  }\;.
$$
It is shown in \cite{FSS} that this always and canonically exists,
it is just the Lie integration $\mathbf{c} = \exp(\mu)$ of the original
$L_\infty$-cocycle.
\footnote{Here and in the following we use $U(1) = \mathbb{R}/\mathbb{Z}$ for brevity,
but in general what appears is $\mathbb{R}/\Gamma$, for $\Gamma\hookrightarrow \mathbb{R}$
the discrete subgroup of \emph{periods} of $\mu$; see \cite{FSS} for details.}

\smallskip
Now, as indicated in the Introduction,
the local Lagrangian for the non-perturbative WZW model is to be an
\emph{$n$-connection} on this $n$-bundle whose curvature $n+1$-form
is $\mu(\theta_{\mathrm{global}})$, the value of the original cocycle
applied to a \emph{globally defined} Maurer-Cartan form on $G$.
Every higher group (in cohesive geometry \cite{cohesive})
does carry a higher Maurer-Cartan form
(see also \cite{hgp}), given by a canonical map
$\theta_G : G \to \flat_{\mathrm{dR}}\mathbf{B}G$
with values in the (nonabelian) \emph{de Rham hypercohomology} stack
$\flat_{\mathrm{dR}}\mathbf{B}G$.
Exactly as $[\omega_G]$ for a Lie group
is represented by the closed left-invariant 3-form $\omega_G=\mu(\theta_G\wedge\theta_G\wedge\theta_G)$, where $\theta_G$ is the Maurer-Cartan form of $G$, the morphism $\mathbf{\Omega}\mathbf{c}$ has a canonical factorization
\[
\xymatrix{
   G
     \ar[drr]_{\mathbf{\Omega}\mathbf{c}}\ar[r]^-{\theta_G}
	 &
	 \flat_{\mathrm{dR}}\mathbf{B}G
	 \ar[r]^-{\flat_{\mathrm{dR}}\mathbf{c}}
	 &
	 \flat_{\mathrm{dR}}
	 \mathbf{B}^{n+1} U(1)\\
&& \mathbf{B}^{n}U(1)\;,
\ar[u]_{\mathrm{curv}}}
\]
where
$\flat_{\mathrm{dR}}\mathbf{B}G$ and $\flat_{\mathrm{dR}}\mathbf{B}^nU(1)$ are the higher smooth stacks of flat $G$-valued and of flat $\mathbf{B}^{n}U(1)$-valued differential forms, respectively, $\theta_G$ is the Maurer-Cartan form, and
$\mathrm{curv}:\mathbf{B}^{n}U(1)\to \flat_{\mathrm{dR}}\mathbf{B}^{n+1} U(1)$ is the canonical curvature morphism (see \cite{FSS,hgp} for details).

\smallskip
There is, however, a fundamental difference between the general case of a higher smooth group and the classical case of a compact Lie group. Namely, the higher Maurer-Cartan form $\theta_G:\mathbf{B}G\to \flat_{\mathrm{dR}}\mathbf{B}G$ will not, in general, be represented by a globally defined flat differential form with coefficients in the $L_\infty$-algebra $\mathfrak{g}$. In other words, we do not have, in general, a factorization
\[
\xymatrix{
& \Omega^1_{\mathrm{flat}}(-;\mathfrak{g})\ar[d]\\
G\ar@{..>}[ru]\ar[r]^{\theta_G}&\flat_{\mathrm{dR}}\mathbf{B}G
}
\]
as in the case of compact Lie groups.
Rather, in general $\theta_G$ is a genuine hyper-cocycle: a collection
of local differential forms on an atlas for $G$, with gauge transformations where their
domain of definition overlaps and higher gauge transformations
on higher intersections.
The universal way to force a globally defined curvature form is
to consider the smooth stack $\tilde{G}$ which is the universal solution to the above factorization problem. That is, we consider the (higher) smooth stack $\tilde{G}$
defined as the following homotopy pullback
\[
\xymatrix{
\tilde{G}\ar[rr]^-{\theta_{\mathrm{global}}} \ar[rr]_<{\rfloor}\ar[d]&& \Omega^1_{\mathrm{flat}}(-;\mathfrak{g})\ar[d]\\
G\ar[rr]^-{\theta_G}&&\flat_{\mathrm{dR}}\mathbf{B}G
}
\]
in higher supergeometric smooth stacks.
In conclusion then the
non-perturbative WZW-model induced by the cocycle $\mu$ is to be
an $n$-connection local Lagrangian of the form
$$
  \mathcal{L}_{\mathrm{WZW}}
  :
  \xymatrix{
    \tilde G \ar[r] & \mathbf{B}^n U(1)_{\mathrm{conn}}\;,
  }
$$
satisfying two conditions:
\begin{enumerate}
  \item its curvature $(n+1)$-form is the evaluation of $\mu$ on the globally
  defined Maurer-Cartan form;
  \item the underlying $n$-bundle is the higher group cocycle $\Omega\mathbf{c}$
  given by Lie integration of $\mu$.
\end{enumerate}
The following proposition now asserts that this indeed exists canonically and
is essentially uniquely.
\begin{proposition}
  On $\tilde G$ there is an essentially unique factorization
  of the globally defined invariant form $\mu(\theta_{\mathrm{global}})$
  through an extended WZW action functional $\mathcal{L}_{\mathrm{WZW}}$
  $$
    \xymatrix{
	  \tilde G
	  \ar@{-->}[dr]_{\mathcal{L}_{\mathrm{WZW}}}
	  \ar[r]^-{\theta_{\mathrm{global}}}
	  &
	  \Omega_{\mathrm{flat}}(-,\mathfrak{g})
	  \ar[r]^{\mu}
	  &
	  \Omega^{n+1}_{\mathrm{cl}}
	  \\
	  & \mathbf{B}^n U(1)_{\mathrm{conn}}
	  \ar[ur]_{F_{(-)}}\;,
	}
  $$
  such that the underlying smooth class $G \to \mathbf{B}^n U(1)$
  is the looping of the exponentiated cocycle $\mathbf{c} = \exp(\mu)$.
  \label{ConstructionOfTheFullWZWTerm}
\end{proposition}
\proof
One considers the smooth stacks
$\flat\mathbf{B}G$ and $\flat\mathbf{B}^{n+1}U(1)$
of  $G$-principal bundles and $U(1)$-principal $(n+1)$-bundles with flat connections, respectively, together with the canonical morphisms
$\flat_{\mathrm{dR}}\mathbf{B}G\to \flat\mathbf{B}G$ and
$\flat_{\mathrm{dR}}\mathbf{B}^{n+1} U(1)\to \flat\mathbf{B}^{n+1} U(1)$ (again, see \cite{FSS,hgp} for definitions). By naturality of these morphisms one has a homotopy commutative diagram
of the form
\[
\xymatrix{
\flat_{\mathrm{dR}}\mathbf{B}G\ar[r]^-{\flat_{\mathrm{dR}}\mathbf{c}}\ar[d]&\flat_{\mathrm{dR}}\mathbf{B}^{n+1}U(1)\ar[d]\\
\flat\mathbf{B}G\ar[r]^-{\flat\mathbf{c}}&\mathbf{B}^{n+1}U(1)\;.
}
\]
Then, by naturally of the inclusions
$\Omega^1_{\mathrm{flat}}(-;\mathfrak{g})\to \flat_{\mathrm{dR}}\mathbf{B}G$ and $\Omega^{n+1}_{\mathrm{cl}}=\Omega^1_{\mathrm{flat}}(-;\mathbb{R}[n])\to \flat_{\mathrm{dR}}
\mathbf{B}^{n+1} U(1)$, one has a homotopy commutative diagram
\[
\xymatrix{
\Omega^1_{\mathrm{flat}}(-;\mathfrak{g})\ar[r]^{\mu}\ar[d]&\Omega^{n+1}_{\mathrm{cl}}\ar[d]\\
\flat_{\mathrm{dR}}\mathbf{B}G\ar[r]^-{\flat_{\mathrm{dR}}\mathbf{c}}&\flat_{\mathrm{dR}}
 \mathbf{B}^{n+1} U(1)\;.
}
\]
Finally, since by definition $\flat_{\mathrm{dR}}\mathbf{B}G$ is the homotopy fiber of the forgetful morphism $
\flat\mathbf{B}G\to \mathbf{B}G$, we have a homotopy pullback diagram of the form
\[
\xymatrix{
\hskip-1emG\simeq \mathbf{\Omega}\mathbf{B}G\ar[r]\ar[d]&\flat_{\mathrm{dR}}\mathbf{B}G\ar[d]\\
\ast\ar[r]&\flat\mathbf{B}G\;.
}
\]
Pasting together the above three diagrams
and the homotopy commutative diagram defining $\tilde{G}$ we obtain the big homotopy commutative diagram
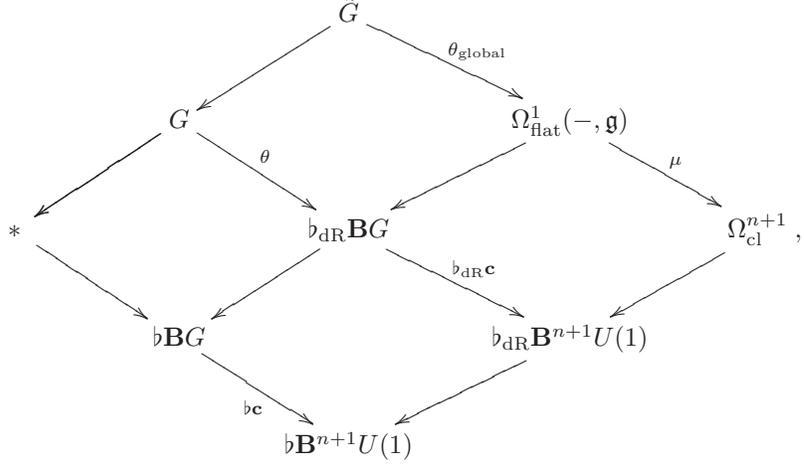
\[
\xymatrix{
	  && \tilde G
	  \ar[dl]
	  \ar[dr]^{\theta_{\mathrm{global}}}
	  \\
	  & G
	  \ar[dl]
	  \ar[dr]^-\theta
	  \ar[dl]
	  &&
	  \Omega^1_{\mathrm{flat}}(-,\mathfrak{g})
	  \ar[dl]
	  \ar[dr]^{\mu}
	  \\
	  {\phantom{mm}}\ast\phantom{mm}
	  \ar[dr]
	  &&  \flat_{\mathrm{dR}}\mathbf{B}G
	  \ar[dl]
	  \ar[dr]^{\flat_{\mathrm{dR}}\mathbf{c}}
	  &&
	  \Omega^{n+1}_{\mathrm{cl}}\;,
	  \ar[dl]
	  \\
	  & \flat \mathbf{B}G
	  \ar[dr]_{\flat \mathbf{c}}
	  &&
	  \flat_{\mathrm{dR}}\mathbf{B}^{n+1}U(1)	  \ar[dl]
	  \\
	  &&
	  \flat \mathbf{B}^{n+1}U(1)
	}
\]
and hence the homotopy commutative diagram
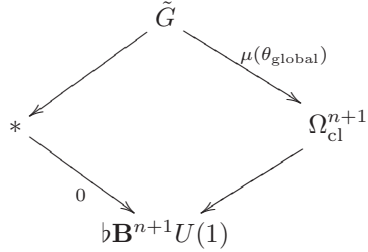
\[
\xymatrix{
	  & \tilde G
	  \ar[dl]
	  \ar[dr]^{\mu(\theta_{\mathrm{global}})}
	  \\
	   \ast
	  \ar[dr]_{0}
	  &&
	  \Omega^{n+1}_{\mathrm{cl}}
	  \ar[dl]
	  \\
	  &  \flat\mathbf{B}^{n+1} U(1)
	 	}
	\,
\]
as the outermost part of the above big diagram. Then, by the universal property of the homotopy pullback, this factors essentially uniquely as
  $$
    \raisebox{49pt}
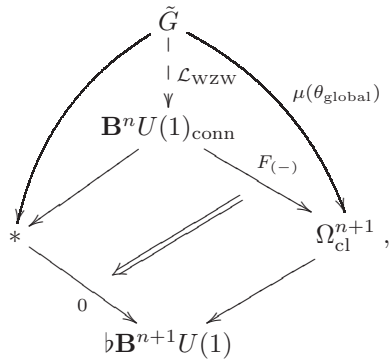
{
    \xymatrix{
	  & \tilde G
	    \ar@/_1pc/[ddl]
	    \ar@/^1pc/[ddr]^{\mu(\theta_{\mathrm{global}})}
		\ar@{-->}[d]^{\mathcal{L}_{\mathrm{WZW}}}
	  \\
	  & \mathbf{B}^n U(1)_{\mathrm{conn}}
	  \ar[dl]
	  \ar[dr]^{F_{(-)}}_{\ }="s"
	  \\
	  \ast
	    \ar[dr]_0^{\ }="t"
	  && \Omega^{n+1}_{\mathrm{cl}}\;,
	  \ar[dl]
	  \\
	  & \flat \mathbf{B}^{n+1}U(1)
	  \ar@{=>} "s"; "t"
	}
	}
	  $$
where we have used the fact that the stack $\mathbf{B}^nU(1)_{\mathrm{conn}}$ of $U(1)$-$n$-bundles with connection is naturally the homotopy fiber of the inclusion $\Omega^{n+1}_{\mathrm{cl}}\to \flat \mathbf{B}^{n+1}U(1)$; see \cite{FSS}.
\endofproof

\begin{remark}
The above proposition has been stated having in mind a cocycle with integral periods, so that $\mathbb{R}/\mathbb{Z}\cong U(1)$. The generalization to an arbitrary subgroup of periods $\Gamma\hookrightarrow \mathbb{R}$ is immediate.
\end{remark}

 \begin{remark}
  The construction of the full higher WZW term
  $\mathcal{L}_{\mathrm{WZW}}$ in Prop. \ref{ConstructionOfTheFullWZWTerm}
  turns out to canonically exhibit the higher WZW-type theory as the boundary theory
  of a higher Chern-Simons-type theory, in the precise sense of
 Def. Prop. \ref{BoundaryConditions}.
  To see this, first recall from \cite{hgp, lpqft, cohesive} that
  an $(n+1)$-dimensional local Chern-Simons-type prequantum field theory
  for a cocycle $\mathbf{c} : \mathbf{B}G \to \mathbf{B}^{n+1}U(1)$
  as above is a map of smooth higher moduli stacks of the form
  $
    \mathcal{L}_{\mathrm{CS}} : \mathbf{B}G_{\mathrm{conn}}
	\to \mathbf{B}^{n+1}U(1)_{\mathrm{conn}}
  $
  which fits into a homotopy commutative diagram of the form
  $$
    \raisebox{30pt}{
    \xymatrix{
	  \flat \mathbf{B}G \ar[rr]^{\flat \mathbf{c}}
	  \ar[d] && \flat\mathbf{B}^{n+1}U(1) \ar[d]
	  \\
	  \mathbf{B}G_{\mathrm{conn}}
	  \ar[d]
	  \ar[rr]^-{\mathcal{L}_{\mathrm{CS}}}
	  &&
	  \mathbf{B}^{n+1}U(1)_{\mathrm{conn}}
	  \ar[d]
	  \\
	  \mathbf{B}G
	  \ar[rr]^-{\mathbf{c}}
	  &&
	  \mathbf{B}^{n+1}U(1)\;.
	}
	}
  $$
This  hence is a refinement to differential cohomology that respects both the inclusion
  of flat higher connections as well as the underlying universal principal
  $n$-bundles. In \cite{FSS} is given a general construction
  of such $\mathcal{L}_{\mathrm{CS}}$ by a stacky/higher version of Chern-Weil theory,
  which applies whenever the cocycle $\mu$ is in transgression with an
  invariant polynomial on the $L_\infty$-algebra $\mathfrak{g}$.
  For instance ordinary 3d Chern-Simons
  theory is induced this way from the transgressive 3-cocycle
  $\langle -,[-,-]\rangle$ on a semisimple Lie algebra, and the
  nonabelian 7d Chern-Simons theory on String 2-connections
  which appears in quantum corrected 11d supergravity is induced
  by the corresponding 7-cocycle \cite{7d}.

  \smallskip
  Now by pasting this 
  diagram
  below
  the diagram
$$
  \raisebox{30pt}{
  \xymatrix{
     \tilde G \ar[r]^-{\theta_{\mathrm{global}}}_<{\rfloor} \ar[d]
	 & \Omega^1_{\mathrm{flat}}(-,\mathfrak{g})
	 \ar[d]
	 \ar[rr]^-{\mu}
	 &&
	 \Omega^{n+1}_{\mathrm{cl}}
	 \ar[d]
	 \\
	 G \ar[r]^-{\theta_G}
	 & \flat_{\mathrm{dR}}\mathbf{B}G
	 \ar[rr]^-{\flat_{\mathrm{dR}} \mathbf{c}}
	 &&
	 \flat_{\mathrm{dR}}\mathbf{B}^{n+1} U(1)
  }
  }
$$
appearing
  in the proof of Prop. \ref{ConstructionOfTheFullWZWTerm}
  we obtain the homotopy commutative diagram
  of smooth higher moduli stacks
  $$
  \raisebox{30pt}{
  \xymatrix{
     \tilde G \ar[r]^-{\theta_{\mathrm{global}}}_<{\rfloor} \ar[d]
	 & \Omega^1_{\mathrm{flat}}(-,\mathfrak{g})
	 \ar[d]
	 \ar[rr]^-{\mu}
	 &&
	 \Omega^{n+1}_{\mathrm{cl}}
	 \ar[d]
	 \\
	 G \ar[r]^-{\theta_G}_<{\rfloor}
	 \ar[d]
	 & \flat_{\mathrm{dR}}\mathbf{B}G
	 \ar[rr]^-{\flat_{\mathrm{dR}} \mathbf{c}}
	 \ar[d]
	 &&
	 \flat_{\mathrm{dR}}\mathbf{B}^{n+1} U(1)
	 \ar[d]
	 \\
	 \ast \ar[r]
	 \ar[d]
	 &
     \flat \mathbf{B}G	
	 \ar[rr]^-{\flat \mathbf{c}}
	 \ar[d]
	 &&
	 \flat \mathbf{B}^{n+1} U(1)
	 \ar[d]
	 \\
	 \ast
	 \ar[r]
	 &
	 \mathbf{B}G_{\mathrm{conn}}
	 \ar[rr]^-{\mathcal{L}_{\mathrm{CS}}}
	 &&
	 \mathbf{B}^{n+1}U(1)_{\mathrm{conn}}\;.
  }
  }
$$
Inside the above diagram 
one reads the
correspondence 
$$
  \raisebox{20pt}
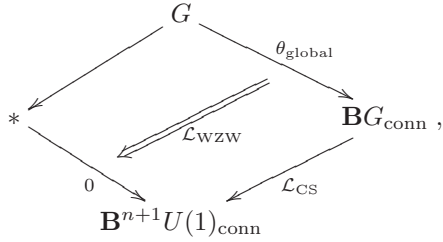
{
  \xymatrix{
    & \tilde G
	\ar[dl]
	\ar[dr]^{\theta_{\mathrm{global}}}_{\ }="s"
	\\
	\ast
	\ar[dr]_0^{\ }="t"
	&&
	\mathbf{B}G_{\mathrm{conn}}\;,
	\ar[dl]^{\mathcal{L}_{\mathrm{CS}}}
	\\
	& \mathbf{B}^{n+1} U(1)_{\mathrm{conn}}
	\ar@{=>}^{\mathcal{L}_{\mathrm{WZW}}} "s"; "t"
  }
  }
$$
which equivalently expresses the higher WZW term as a cocycle in
degree $n$ differential cohomology twisted by the
Chern-Simons term 
evaluated on the
globally defined Maurer-Cartan
form.
  According to definition \ref{BoundaryConditions} this precisely exhibits $\mathcal{L}_{\mathrm{WZW}}$
as a boundary condition for $\mathcal{L}_{\mathrm{CS}}$.

 \smallskip
  This general mathematical statement seems to be well in
  line with the relation between higher Chern-Simons terms and
  higher WZW models found in \cite{Witten}.
  Notice that with $\mathcal{L}_{\mathrm{WZW}}$ realized as a boundary theory
  of $\mathcal{L}_{\mathrm{CS}}$
  this way, any further boundary of $\mathcal{L}_{\mathrm{WZW}}$,
  notably as in Def. \ref{BoundaryCondition}, makes that a \emph{corner}
  of $\mathcal{L}_{\mathrm{CS}}$. In fact, in \cite{lpqft} is
  shown that $\mathcal{L}_{\mathrm{CS}}$ itself is already naturally
  a boundary theory for a topological field theory of yet one dimension
  more, namely a universal higher topological Yang-Mills theory.
  Hence we find here a whole cascade of \emph{corner field theories}
  of arbitrary codimension. For instance from the results above we have
  the sequence of higher order corner theories that looks like
 \vskip 2mm
 \hskip-0.8cm \scalebox{0.82}{
  $$
    \xymatrix{
	  \mbox{M2-brane}
	  ~\ar@{^{(}->}[rr]^-{\mbox{\tiny ends on}} &&
	  \mbox{M5-brane}
	  ~\ar@{^{(}->}[rrr]^-{\mbox{\tiny WZW boundary of}} &&& \mbox{7d CS in 11d Sugra}
	  ~\ar@{^{(}->}[rr]^-{\mbox{\tiny boundary of}}
	  &&
	  \mbox{8d tYM}
	}
	\,.
  $$}

  \vskip 3mm
  \noindent
  Such hierarchies of higher order corner field theories have
  previously been recognized and amplified in string theory and M-theory
  \cite{SatiC, boundary, F}. More discussion of the above formalization of these
  hierarchies in local (multi-tiered) prequantum field theory is in \cite{lpqft}.
  Closely related considerations have appeared in \cite{Freed432}.
\end{remark}
To further appreciate the abstract construction of the higher
WZW term $\mathcal{L}_{\mathrm{WZW}}$ in Prop. \ref{ConstructionOfTheFullWZWTerm},
it is helpful
to notice the following two basic examples, which are in a way
at opposites ends of the space of all examples.
\begin{example}
  For $\mathfrak{g}$ an ordinary (super-)Lie algebra and $G$ an ordinary
  (super-)Lie group integrating it, we have
  $\flat_{\mathrm{dR}}\mathbf{B}G \simeq \Omega^1_{\mathrm{flat}}(-,\mathfrak{g})$
  \cite{cohesive}. This implies that in this case $\tilde G \simeq G$,
  hence that there is no extra ``differential extension''.
  Now for $\mu$ a 3-cocycle, the induced $\mathcal{L}_{\mathrm{WZW}}$
  is the traditional WZW term, refined to a Deligne 2-cocycle/bundle gerbe
  with connection as in \cite{Ga, FW}.
  \label{FullWZWOnLieGroup}
\end{example}
\begin{example}
  For $\mathfrak{g} = \mathbb{R}[n]$ we can take the smooth higher group
  integrating it to be the $(n+1)$-group
  $G = \mathbf{B}^nU(1)$.
  In this case, as shown in \cite{cohesive}, the definition of $\tilde G$ is precisely the
  characterization of the moduli $n$-stack of $U(1)$-$n$-bundles with connections, so that
  $
    \tilde G \simeq \mathbf{B}^n U(1)_{\mathrm{conn}}
	\,
  $
  in this case.
  Then for $\mu : \mathfrak{g} \to \mathbb{R}[n]$
  the canonical cocycle (the identity), it follows that
  $\mathcal{L}_{\mathrm{WZW}}$ is the identity, hence is the
  canonical $U(1)$-$n$-connection on the moduli $n$-stack of all
  $U(1)$-$n$-connections.
  This describes the extreme case of a higher WZW-type field theory with
  \emph{no} $\sigma$-model fields and \emph{only} a ``tensor field''
  on its worldvolume, and whose action functional is simply the
  higher volume holonomy of that higher gauge field.
  \label{FullWZWOnCirclenGroup}
\end{example}
Generic examples of higher WZW theories are twisted products
of the above two basic examples:
\begin{example}
  Consider $K$ a higher (super-)group extension of a Lie (super-)group $G$ of the form
  $
    \xymatrix{
	  \mathbf{B}^n U(1)
	  \ar[r]
	  &
	  K
	  \ar[r]
	  &
	  G
	}
	\,.
  $
  For instance $G$ may be a translation super-group $\mathbb{R}^{d;N}$
  and $K$ the Lie integration of one of the extended superspaces
  such as $\mathfrak{m}2\mathfrak{brane}$ considered above
  (spacetime filled with a brane condensate, Remark \ref{BraneCondensates}).
  This means that $K$ is a \emph{twisted product}  of
  the (super-)Lie group $G$ and the $(n+1)$-group $\mathbf{B}^n U(1)$,
  which appear in examples \ref{FullWZWOnLieGroup} and
  \ref{FullWZWOnCirclenGroup} above. Since the construction of
  $\mathcal{L}_{\mathrm{WZW}}$ in the proof of Prop. \ref{ConstructionOfTheFullWZWTerm}
  suitably respects products, it follows that
  the field content of a higher WZW model on the higher smooth (super-)group $K$ is
  a tuple consisting of
  \begin{enumerate}
    \item a $\sigma$-model field with values in $G$;
	\item an $n$-form higher gauge field,
  \end{enumerate}
  both subject to a twisting condition which gives the
  higher gauge field a twisted Bianchi identity depending on the
  $\sigma$-model fields.

  \smallskip
  In particular, for the extended spacetime given by an M2-brane condensate in
  11-dimensional $(N=1)$-super spacetime, this says that the M5-brane
  higher WZW model according to Section \ref{The5brane}
  has fields given by a multiplet consisting of embedding fields into
  spacetime and a 2-form higher gauge field (``tensor field'') on its
  worldvolume. Notice that the higher gauge transformations of the 2-form
  field are correctly taken into account by this
  full (in particular non-perturbative) construction of the WZW term
  as a higher prequantum bundle.
\end{example}

\smallskip

\noindent{\bf Acknowledgements.}
We thank Friedemann Brandt for helpful discussion and for corrections.
U.S. also thanks Joost Nuiten for useful discussion.
D.F. thanks the University of Pittsburgh for invitation in summer 2013.
The research of H.S. is supported by NSF Grant PHY-1102218.
U.S. thanks the University of Pittsburgh for invitations in spring
2013 and again in summer 2013, during which this work was completed.

\end{document}